\documentclass[10pt]{article}
\usepackage{amsmath,amsfonts,amssymb,amsthm,amscd,accents,mathtools}
\usepackage{epsfig,url,cite,graphicx,graphics,subfigure,psfrag}
\usepackage{vmargin} 
\usepackage{fancyhdr}
\newcommand{\B}[1]{\mathbf{#1}} 
\newcommand{\norm}[1]{\lVert#1\rVert}
\newcommand{\real}{\mathbb{R}}
\newcommand{\mwcphi}{\B{\Phi}}
\newcommand{\mwcpsi}{\B{\Psi}}
\newcommand{\rdphi}{\B{\Sigma}}
\newcommand{\rdpsi}{\B{\Psi}}

\newcommand{\rect}[1]{\text{rect}\bigl(#1\bigr)}
\newcommand{\sinc}[1]{\text{sinc}#1}
\newcommand{\single}{\renewcommand{\baselinestretch}{1}\large\normalsize}
\newcommand{\double}{\renewcommand{\baselinestretch}{1.4}\large\normalsize}
\newcommand{\mwcM}{M'}
\newcommand{\mwcL}{L'}
\newcommand{\mwcq}{q'}
\newcommand{\mwcW}{W'}

\setpapersize{A4}
\setmargins{1in}{.75in}{6.25in}{9in}{12pt}{10pt}{20pt}{30pt}
\title{\vspace*{-20pt}\textbf{Reconciling Compressive Sampling Systems for Spectrally-sparse Continuous-time Signals}}
\author{Michael~A.~Lexa, Mike~E.~Davies and John~S.~Thompson\\ \url{{michael.lexa, mike.davies, john.thompson}@ed.ac.uk}\\ \\
Institute of Digital Communications\\
University of Edinburgh}
\date{Jan 2011, Revised May 2011, Revised again Aug 2011}

\begin{document}
\sloppy\double
\maketitle

\begin{abstract}
The Random Demodulator (RD) and the Modulated Wideband Converter (MWC) are two recently proposed compressed sensing (CS) techniques for the acquisition of continuous-time spectrally-sparse signals.
They extend the standard CS paradigm from sampling discrete, finite dimensional signals to sampling continuous and possibly infinite dimensional ones, and thus establish the ability to capture these signals at sub-Nyquist sampling rates.
The RD and the MWC have remarkably similar structures (similar block diagrams), but their reconstruction algorithms and signal models strongly differ.
To date, few results exist that compare these systems, and owing to the potential impacts they could have on spectral estimation in applications like electromagnetic scanning and cognitive radio, we more fully investigate their relationship in this paper.
We show that the RD and the MWC are both based on the general concept of random filtering, but employ significantly different sampling functions.
We also investigate system sensitivities (or robustness) to sparse signal model assumptions.
Lastly, we show that ``block convolution'' is a fundamental aspect of the MWC, allowing it to successfully sample and reconstruct block-sparse (multiband) signals.
Based on this concept, we propose a new acquisition system for continuous-time signals whose amplitudes are block sparse.
The paper includes detailed time and frequency domain analyses of the RD and the MWC that differ, sometimes substantially, from published results.
\end{abstract}

\section{Introduction}\label{sect:introduction}
The theory of compressed sensing (CS) says that if a signal is sufficiently sparse with respect to some basis or frame, it can be faithfully reconstructed from a small set of linear, nonadaptive measurements~\cite{donoho2006,candes2006,candes_wakin2008}.
When the signal belongs to a finite dimensional space, this statement means that it can be reconstructed from a set of measurements whose cardinality may be significantly less than the space's dimension.
It also implies that the measurement process is described by an underdetermined linear system of equations, or equivalently, by a rectangular matrix with more columns than rows.
The fundamental work of Cand\'{e}s, Romberg, and Tao~\cite{candes_romberg_tao2006} and Donoho~\cite{donoho2006} established sufficient conditions upon such sensing matrices, that if satisfied, allow the stable inversion of the linear systems.
A key aspect of CS, and one which plays an important role in this paper, is that sensing matrices drawn at random%
\footnote{There are several ways to construct viable random sensing matrices. For example, its entries could simply be independent and identically distributed realisations of a a zero mean, unit variance Gaussian random variable.}
often satisfy these conditions.

Conceptually, CS theory has three main thrusts: (1) the development of recovery methods that efficiently and faithfully reconstruct the original signal from its compressed samples, (2) the investigation of new signal models that effectively represent signal sparsity or other signal structure, and (3) the creation of new sampling (measurement) mechanisms that acquire signals in a compressed manner.
The first concerns the reconstruction process and asks how one specifically reconstructs the original signal from the CS measurements (see, e.g.,\cite{donoho2006,candes_romberg_tao2006,tropp2007}).
The second concerns the examination of different signal classes of interest and asks if are structured representations that can be exploited~\cite{baraniuk_etal2010,eldar2009,lu_do2008}.
The third concerns the design of the physical sampling system and asks how one devises a system to acquire CS measurements~\cite{eldar2009,romberg2009,tropp_etal2010,mishali_eldar2010}.
This paper focuses on the third thrust and examines two sampling systems for two distinct, but related, signal models. 
 
Several CS based signal acquisition systems have been proposed for both continuous (analogue) and discrete signals.
For example, the single-pixel camera~\cite{duarte_etal2008} is a novel compressive imaging system, where light is projected onto a random basis using a micro-mirror device, and then the projected image is captured by a single photo-diode (the single ``pixel'').
Other examples include random filtering~\cite{tropp_wakin_etal2006} and random convolution~\cite{romberg2009} that advocate random linear filtering and low rate sampling as a means to collect CS measurements.
In these cases, ``random'' filters are linear filters whose impulse responses are realisations of particular random processes.
Along the same lines, the \emph{Random Demodulator} (RD)~\cite{kirolos_etal2006,laska_etal2007,tropp_etal2010} and the \emph{Modulated Wideband Converter} (MWC)~\cite{mishali_eldar2010,mishali_eldar_dcc2010,chen_etal2010,mishali_elron_eldar_icassp2010} have recently been proposed as CS sampling systems that target continuous-time spectrally-sparse signals. 
The RD is a single channel%
\footnote{An early multi-channel random demodulator was proposed in~\cite{yu_hoyos_sadler2008}.}, 
uniform sub-Nyquist sampling strategy for acquiring \emph{sparse multitone signals}; the MWC is a multi-channel, uniform sub-Nyquist sampling strategy for acquiring \emph{sparse multiband signals}.
 (Precise definitions for these two signal classes are provided in Section~\ref{sect:signal_models}.)
The RD and the MWC have tremendous potential impact because of the longstanding, proven usefulness of spectral signal models in many engineering and scientific applications (e.g. communications, radar/sonar, medical imaging, etc.).
Perhaps owing to the near coincidental emergence of these systems, few results exist to date that reconcile their remarkably similar structures (see Figure~\ref{fig:sampling_systems}) with their different reconstruction algorithms.  
In fact, the current literature paints a somewhat artificial dividing line between the RD and the MWC, preferring to focus primarily on one scheme or another rather than drawing connections between them.
One exception is the recent comparative analysis by Mishali et al.~\cite{mishali_eldar_elron2011} that focuses on the systems' robustness to signal model perturbations and computational and hardware complexity.  
We comment more on~\cite{mishali_eldar_elron2011} below.
 
In this paper, we offer new insights into the relationship between the RD and the MWC that complement the original works of Tropp et al.~\cite{tropp_etal2010} and Mishali and Eldar~\cite{mishali_eldar2010}. 
We apply tools from modern sampling theory and classical Fourier analysis and show that the RD and the MWC are two manifestations of the same CS sampling approach, namely random filtering/convolution~\cite{tropp_wakin_etal2006,romberg2009}. 
This fact is reflected the systems' similar structure.
At the same time, we show that the sampling functions characterising the systems strongly distinguish the two schemes.
In Section~\ref{sect:signal_models}, we examine three different properties of the RD and the MWC related to the underlying assumption on signal sparsity.
We discuss how sparsity manifests itself in each case and comment on the system's sensitivity or robustness to changes in sparsity.
In Section~\ref{sect:more_models}, we highlight, among other insights, the MWC's use of block convolution as a principal processing step that enables it to successfully sample and recover ``block-sparse'' signals, i.e. signals whose nonzero components are grouped together.
Extending this idea, we propose a new CS based sampling system and show through an example that it can successfully sample and reconstruct continuous-time signals that are block sparse in the time domain.

To be clear, we do not discuss the conditions of successful reconstruction, nor implementation issues in this paper.
The original works of Tropp et al.~\cite{tropp_etal2010} and Mishali and Eldar~\cite{mishali_eldar2010}, and even some subsequent scholarship~\cite{chen_etal2010,mishali_elron_eldar_icassp2010,mishali_eldar_dcc2010}, extensively investigate these issues.
Some of the reconstruction conditions will be stated in the descriptions of the systems in Section~\ref{sect:sampling_mechanisms}, but the presumption throughout the paper is that the RD and the MWC are theoretically proven CS based techniques to sample and reconstruct continuous-time spectrally-sparse signals.
In addition, we only examine the idealised RD and MWC because their fundamental similarities and differences are sharper in this setting (e.g., one does not have to account for the effects of non-ideal filters).
Aspects of practical implementations are discussed in~\cite{kirolos_etal2006} and~\cite{mishali_etal2011}.

This paper and the comparative analysis presented in~\cite{mishali_eldar_elron2011} are similar in some respects; however, the approach and the conclusions are very different.
For example, both touch on a model sensitivity issue of the RD, but in~\cite{mishali_eldar_elron2011} this sensitivity is billed as a fundamental shortcoming in comparison to the MWC because it does not exhibit the exact same sensitivity.
In contrast, we argue here that the sensitivity is a manifestation of a CS sensitivity and that the MWC also inherits a shortcoming from CS theory, albeit a different one than the RD.
In short, the approach of the present paper asks if there is an underlying link in the way the RD and the MWC process their signals and then examines their similarities and differences from this common perspective.
This perspective yields a consistent and broader understanding of these systems and their relation to other CS schemes and standard sampling theory.

Throughout the paper, we denote time domain signals by lower case letters (e.g. $x,y,\psi$) and frequency domain signals by upper case letters (e.g. $X,Y$).
Vectors and matrices are indicated by boldface type (e.g. $\B{x},\B{Y},\B{\Phi}$).
Parameters are denoted by upper case letters with one exception: the number of channels in the MWC is denoted by $\mwcq$.

\textbf{Contributions.}
The main contributions of this paper are:
\begin{itemize}
\item a consistent analysis that (i) clearly shows random filtering underlies the RD and the MWC and (ii) highlights system sensitivities
\item the insight that block convolution fundamentally enables the MWC to sample and recover frequency block-sparse signals, and the generalization of this idea to a new sampling system.
\end{itemize}

\section{Sampling mechanisms and signal models}\label{sect:sampling_mechanisms}
In this section, we examine the sampling mechanisms of the RD and the MWC from a modern sampling theory perspective.
We show the output samples for both systems are equal to the inner products of the input signal with a set of sampling functions that arise from the systems' designs.
We observe that unlike typical sampling functions, these sampling functions involve random waveforms, a central component in many CS sampling systems. 
If the inner products are interpreted as analogue filtering operations, we show that the samples result from a generalised random filtering or random convolution as described by Romberg~\cite{romberg2009} and Tropp et al.~\cite{tropp_wakin_etal2006} as a means to acquire CS samples.
This analysis suggests that the RD and the MWC are two manifestations of the same sampling approach, but differ in the specific form of the sampling functions.
The difference in sampling functions also reflects the difference in the assumed signal models for the RD and the MWC.
We do not introduce the notion of signal sparsity in this section because the conclusions reached do not depend on this aspect.
Signal sparsity and its consequences are discussed in Section~\ref{sect:signal_models}.

\subsection{Sampling with the random demodulator}\label{subsect:sampling_with_rd}
\begin{figure}
\subfigure[Random demodulator]{
\begin{minipage}{0.48\textwidth}
\centerline{\includegraphics[width=5.5cm]{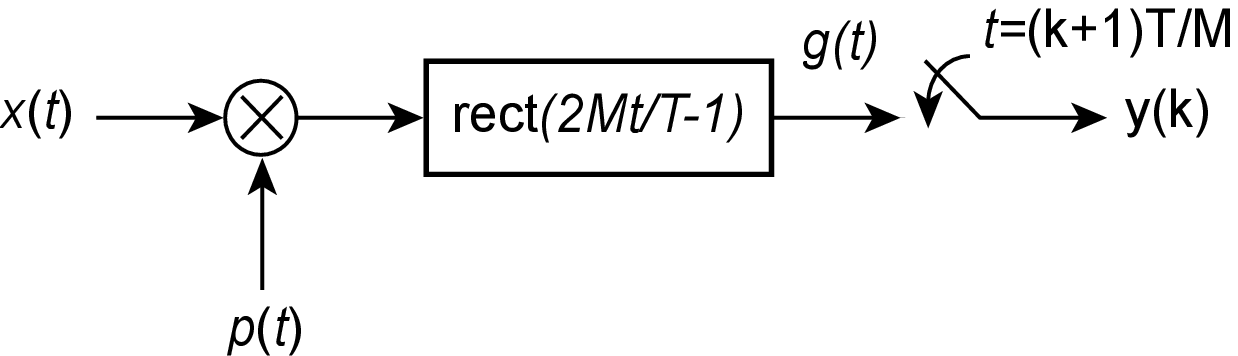}}
\end{minipage}\label{subfig:rd}}
\subfigure[Modulated wideband converter]{
\begin{minipage}{0.48\textwidth}
\centerline{\includegraphics[width=6.5cm]{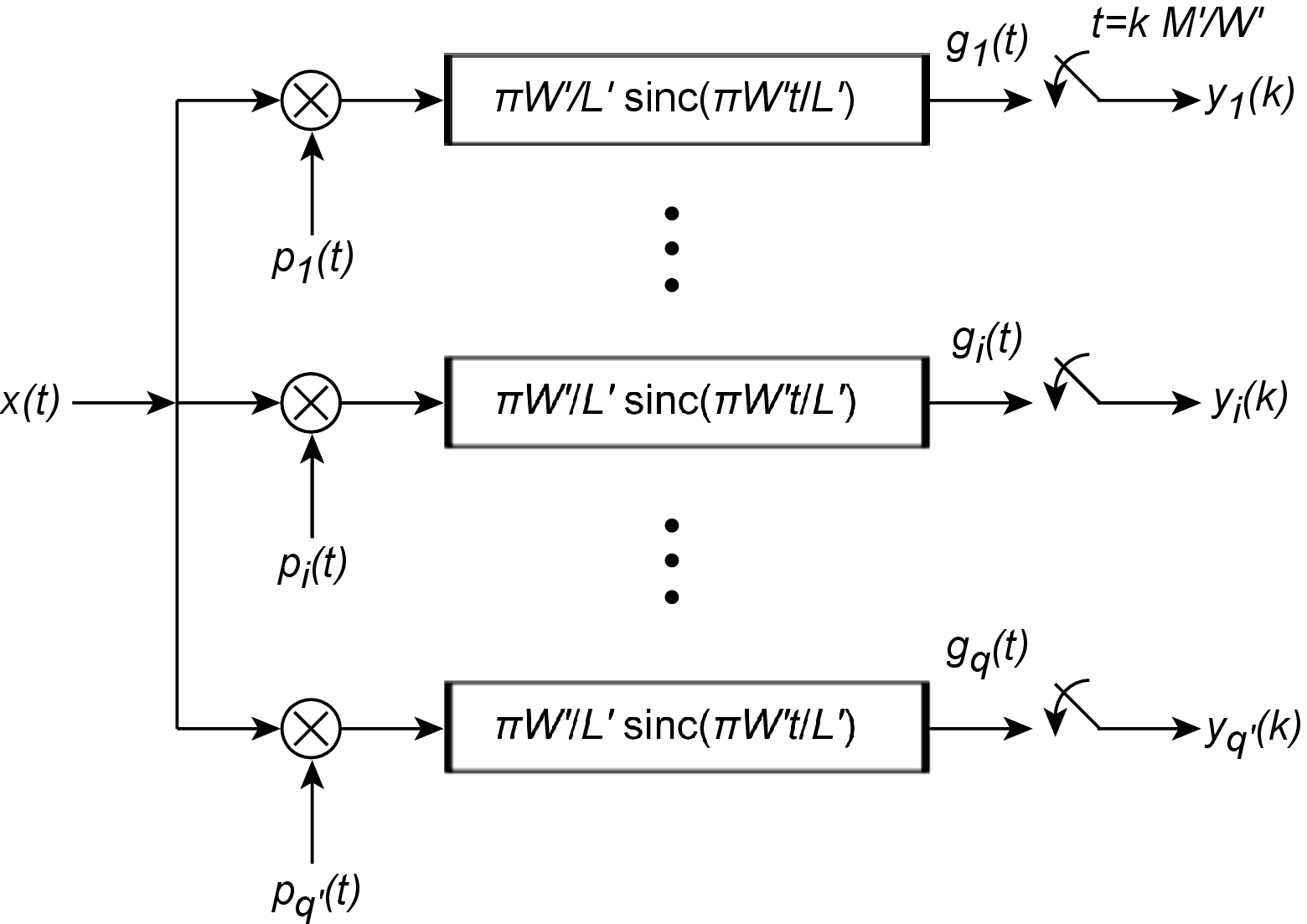}}
\end{minipage}\label{subfig:mwc}}
\caption{\small Time domain block diagrams of the random demodulator (RD) and the modulated wideband converter (MWC).
The RD is characterised by $T$, the duration of the observation interval and $M$, a sampling rate parameter.
The MWC is characterised by $\mwcL$, a parameter for the period of $p_{i}(t)$, $\mwcM$, a sampling rate parameter, and $\mwcq$, the number of channels.
The primary structural difference between the systems is the type of filter employed prior to sampling---the RD uses an ideal integrator and the MWC uses ideal low pass filters.} \label{fig:sampling_systems}
\end{figure}

Let $x(t)$ be a continuous-time, complex-valued signal defined on the real line.
The RD acquires samples of $x(t)$ on a finite observation interval where here, we assume, without loss of generality, that the samples are collected in the interval $[0,T]$ seconds.
In~\cite{tropp_etal2010}, Tropp et al. adopt a particular signal model for $x(t)$ on this interval.
They assume in part that $x(t)$ has a Fourier series (FS) expansion on $[0,T]$ which has bounded harmonics, i.e., $-W/2\leq \frac{n}{T}<W/2$ Hz for $n\in\mathbb{Z}$.
On this interval, $x(t)$ is therefore modeled as
\begin{equation}\label{equ:multitone_model}
x(t)=\sum_{n=-N/2}^{N/2-1} X(n) e^{j\frac{2\pi}{T}nt}, \quad t\in[0,T],
\end{equation}
where $\{X(n)\}$ denotes the FS coefficients of $x(t)$ and $N=TW$.
For ease of exposition, $N$ is assumed to be an even positive integer.
This signal model is often called a \emph{multitone} model.

To acquire the samples, a RD first multiplies $x(t)$ by a waveform $p(t)$ and then filters and samples the product $x(t)p(t)$ on $[0,T]$ (see Figure~\ref{subfig:rd}).
The signal $p(t)$ is taken to be a realisation of a continuous random process derived from a vector of Bernoulli random variables.
Let $\B{Z}=[Z_0,\dotsc,Z_{L-1}]$ be a vector of independent and identically distributed Bernoulli random variables $Z_{l}$ taking values $\pm1$ with equal probability%
\footnote{A $\pm1$ symmetric Bernoulli random variable is also known as a Rademacher random variable.}
and let $p(t;\B{Z})$ denote the random process
\begin{equation}
p(t;\B{Z})=Z_{l},\quad  t\in \biggl[  \frac{l}{W},  \frac{l+1}{W}\biggr), \quad l=0,\dotsc, N-1.
\end{equation}
A realisation $\B{Z}_{0}$ of $\B{Z}$ produces a single realisation $p(t;\B{Z}_0)$ of $p(t;\B{Z})$.
Here, we abbreviate $p(t;\B{Z}_0)$ by $p(t)$ and thus consider $p(t)$ to be a deterministic quantity, although its randomness plays an important role in proving performance guarantees~\cite{tropp_etal2010}.
In this paper, we sometimes refer to $p(t)$ as a random waveform in deference to this point.
We stress that when acquiring samples on $[0,T]$, the RD uses a single realisation of $p(t;\B{Z})$, but different realisations may be used for other observation intervals.  
Note also that $p(t)$ has the FS representation,
\begin{equation}
p(t)=\sum_{n=-\infty}^{\infty} P(n) e^{j\frac{2\pi}{T}nt}, \quad t\in[0,T]
\end{equation}
where $\{P(n)\}$ is the set of FS coefficients of $p(t)$.

The analogue filter in the RD design is taken to be an ideal integrator with impulse response $h(t)=\text{rect}\bigl(\frac{2M}{T}t-1\bigr)$, where
\begin{equation}\label{equ:rect}
\text{rect}(x)=\begin{cases}
1\quad\text{for~}-1\leq x \leq 1\\
0\quad\text{otherwise}\end{cases},
\end{equation}
and $M\in\mathbb{Z}^{+}$.
The sampling period $T_s$ is taken to be $M$ times shorter than the observation window ($T_s=T/M$).
The system therefore samples at the rate of $M/T$ Hz.
The multitone signal model and the RD sampling system are therefore parameterised by $N$, the parameter equal to the time-frequency product $TW$ and $M$, the parameter that controls the RD's sampling rate.
Here, we assume that $M<N$.

The goal of the RD is to sample $x(t)$ at low rates while retaining the ability to reconstruct it in the interval $[0,T]$.
Reconstruction entails the discovery of the active frequencies (the signal's spectral support) and the amplitude of the corresponding FS coefficients.
If $x(t)$ is spectrally sparse on $[0,T]$, then reconstruction is possible using CS algorithms~\cite{tropp_etal2010}.
In this case, we note that signal reconstruction only implies the recovery of the spectral content of $x(t)$ in the observation interval.
In other words, the samples $y(k), k=0,\dotsc,M-1$, do not convey information about the spectral content of $x(t)$ outside of this interval.
To obtain spectral information outside of $[0,T]$, the RD must be applied to other intervals (of possibly different durations).
If the RD is applied to consecutive intervals, a time-frequency decomposition of $x(t)$ similar to the short-time Fourier transform can be obtained for multitone signals. 

\textbf{Time domain description.} 
By inspection of Figure~\ref{subfig:rd}, the output samples $y(k)$ can be expressed as
\begin{equation}\label{equ:rd_timedomain}
\begin{split}
y(k)&=g\bigl((k+1)\tfrac{T}{M}\bigr) \\
&= \int_{0}^{T}x(\tau)p(\tau)\rect{\tfrac{2M}{T}(t-\tau)-1}~d\tau \Biggr\rvert_{t=(k+1)\tfrac{T}{M}}
\end{split}
\end{equation}
for $k=0,\dotsc,M-1$ where $g(t)=x(t)p(t)*h(t)$ ($*$ denotes convolution).
By substituting~\eqref{equ:multitone_model} into this expression and evaluating the integral, the following equation relating the time domain samples $y(k)$ to the FS coefficients $X(n)$ results:
\begin{equation}\label{equ:rd_linearsystem}
y(k)=  \sum_{n=-N/2}^{N/2-1} \alpha(n) \,X(n) \sum_{l=k\tfrac{N}{M}}^{(k+1)\tfrac{N}{M}-1} p_{l} \, e^{j\tfrac{2\pi}{N}nl},
\end{equation}
for $k=0,\dotsc M-1$ where $p_{l}=p(l/W)$ and 
\begin{equation*}
\alpha(n)=\begin{cases} T \dfrac{e^{j\frac{2\pi}{N}n}-1}{j2\pi n} & n\neq 0 \\
1/W  & n=0 \end{cases}.
\end{equation*}  
Tropp et al. derived~\eqref{equ:rd_linearsystem} in~\cite{tropp_etal2010} by analysing an equivalent digital system. 
In Appendix~\ref{app:RD}, we provide an alternate derivation that explicitly shows the analogue processing inherent in sampling with the RD.  

\begin{figure}
\begin{minipage}{0.48\textwidth}
\centerline{\includegraphics[width=5cm]{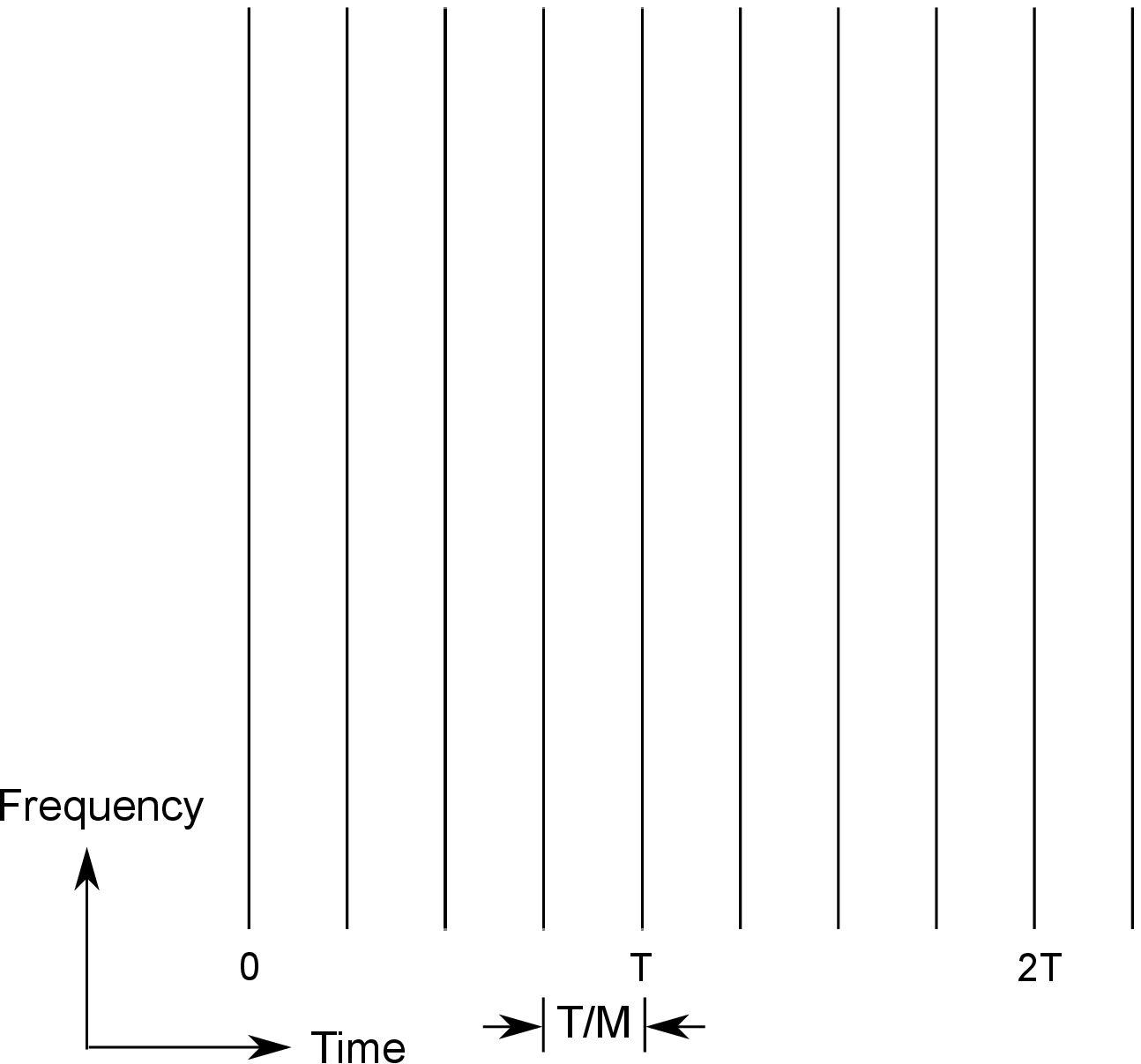}}
\end{minipage}
\begin{minipage}{0.48\textwidth}
\centerline{\includegraphics[width=5cm]{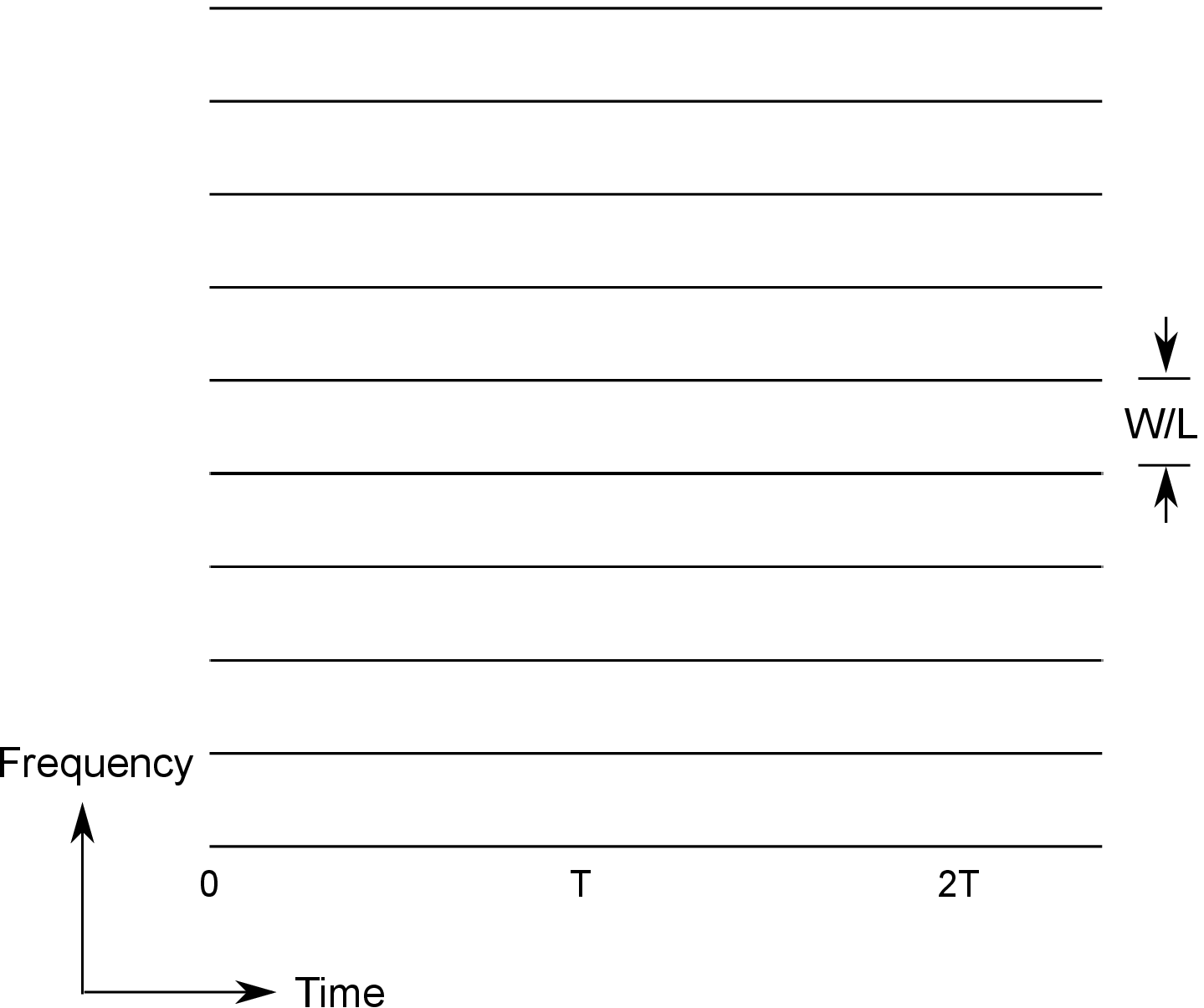}}
\end{minipage}
\caption{\small The output samples of both the RD and the MWC can be described as inner products of the input signal $x(t)$ with certain sets of sampling functions.
The panel on the left depicts the time-frequency support of the RD sampling functions where each vertical strip represents the support of one sampling function.
Similarly, the panel on the right depicts the support of the MWC sampling functions where each horizontal strip represents the support of one sampling function.
For the RD and the MWC, the support characteristics of the sampling functions directly derive from the type of analogue filters used prior to sampling.
The RD and the MWC represent two extreme cases: The RD has perfectly localised support in time but completely unlocalized support in frequency.
The MWC is the exact opposite.}\label{fig:time-freq_support}
\end{figure}

Because sampling is a linear operation with the RD, the samples $y(k)$ can be viewed as inner products of $x(t)$ with the set of sampling functions ${\bigl\{p(\tau)\rect{2k+1-\tfrac{2M}{T}\tau}\bigr\}}$ where
\begin{equation}\label{equ:rd_samplingfunctions}
y(k)= \bigl\langle x(\tau),p(\tau)\text{rect}\bigl(2k+1-\tfrac{2M}{T}\tau\bigr)\bigr\rangle,
\end{equation}
for $k=0,\dotsc,M-1$ and
\begin{equation*}
\bigl\langle x(t),s(t) \bigr\rangle = \int_{0}^{T} x(t)s^{*}(t)~dt,
\end{equation*}
for two continuous functions $x(t)$, $s(t)$ on $[0,T]$.
These sampling functions have finite duration in time ($T/M$ seconds), but because their Fourier transforms involve $\sinc{}$ functions, they extend infinitely in frequency.
In the time-frequency plane, their support partitions the space into vertical strips of equal width (see Figure~\ref{fig:time-freq_support}, left panel).
We note that unlike modern sampling theory~\cite{unser2000}, the sampling functions in~\eqref{equ:rd_samplingfunctions} contain the random waveform $p(t)$, and the conditions they must satisfy to ensure stable recovery is governed by CS theory and not Shannon-Nyquist based sampling theory.
(Refer to~\cite{unser2000} and~\cite{eldar2009} for details regarding the conditions that sampling functions typically must satisfy.)
  
From~\eqref{equ:rd_timedomain}, it is clear the samples $y(k)$ can be thought of as pointwise evaluations of the convolution between $x(t)p(t)$ and an ideal integrator.
Equally valid, however,  is the view that the samples are pointwise evaluations of a random, linear filtering operation involving $x(t)$ and the time-varying analogue filter $h(t,\tau)=p(\tau)\rect{\tfrac{2M}{T}(t-\tau)-1}$,
\begin{equation}\label{equ:rd_randomfiltering}
y(k)=\int_{0}^{T} x(\tau)h(t,\tau)~d\tau \Biggr\rvert_{t=(k+1)\tfrac{T}{M}}
\end{equation} 
for $k=0,\dotsc,M-1$.
Here, the impulse response $h(t,\tau)$ is considered random because at each time instance it is a windowed portion of a signal that randomly alternates between $\pm 1$.
The samples $y(k)$ can therefore be thought of as the result of a random filtering operation, conceptually similar to the random filtering schemes proposed in~\cite{romberg2009} and~\cite{tropp_wakin_etal2006}.
In~\cite{tropp_wakin_etal2006}, Tropp et al. proposed a CS sampling scheme where a sparse discrete-time signal is first filtered by a digital filter whose impulse response is a realisation of a sequence of independent and identically distributed random variables, and then subsampled at a low rate.
They illustrated through examples that with the use of CS recovery algorithms random filtering is a potential sampling structure to acquire CS measurements for sparse discrete time signals.
In~\cite{romberg2009}, Romberg proposed and examined a similar idea but considered a specific digital filter that randomly changes the phase of the input signal.
Interestingly, Romberg considered the RD as a separate, follow-on processing step to his approach instead of considering it as a generalisation to his notion of random convolution.
Here,~\eqref{equ:rd_randomfiltering} shows that the sampling mechanism of the RD can be viewed as a random filtering operation applied to continuous-time signals. 
We note, however, that the filtering operation in~\eqref{equ:rd_randomfiltering} is not a convolution because of the time-varying nature of $h(t,\tau)$.
Strictly speaking then~\eqref{equ:rd_randomfiltering} is distinct from the systems proposed in~\cite{tropp_wakin_etal2006} and~\cite{romberg2009}, although random filtering remains a common thread.

\textbf{Frequency domain description.} 
An equivalent frequency domain expression to~\eqref{equ:rd_timedomain} can be derived (see Appendix~\ref{app:RD}) that relates the discrete Fourier transform (DFT) of $y(k)$, denoted by $Y(n)$, to the FS coefficients $X(n)$,
\begin{equation}\label{equ:rd_freqdomain}
Y(n)= T \sum_{m=n-\tfrac{N}{2}+1}^{n+\tfrac{N}{2}} P(m) e^{-j\frac{2\pi}{T}n}\sinc{(\tfrac{\pi}{M}n)}  \,X(n-m),
\end{equation}
$n=0,\dotsc,M-1$, where $\sinc{(x)}=\sin{(x)}/x, x\in \real$.
This equation clearly shows the frequency domain convolution caused by the multiplication with $p(t)$ and the effect of filtering with an ideal integrator (indicated by the presence of the $e^{-j\frac{2\pi}{T}n}\sinc{(\tfrac{\pi}{M}n)}$ term).
Thus, one can also interpret $Y(n)$ as the output of a random, frequency-varying filter with impulse response $H(n,m)=P(m) e^{-j\frac{2\pi}{T}n}\sinc{(\tfrac{\pi}{M}n)}$.
We see therefore that the RD's output in either the time or frequency domains can be viewed as the output of a random filter or convolution.

\subsection{Sampling with the modulated wideband converter}\label{subsect:mwc}
We now let $x(t)$ be a bandlimited, finite energy signal.
The spectral content of $x(t)$ on $\mathbb{R}$ is then appropriately given by its Fourier transform (FT) $X(\omega)$,
\begin{equation*}
X(\omega)= \int_{-\infty}^{\infty}x(t)\,e^{j\omega t}~dt.
\end{equation*}
Here, $x(t)$ is bandlimited in the usual sense, i.e., $X(\omega)$ is assumed to be bounded:  $X(\omega)=0$ for $\lvert \omega\rvert \geq \pi \mwcW$ radians per second, $\mwcW\in\mathbb{R}^{+}$, where $\pi \mwcW$ is the bandwidth of $x(t)$ and $2\pi \mwcW$ is the Nyquist frequency in radians per second.
We adopt the following definition from~\cite{bresler2008}.
The class of \emph{multiband} signals $\mathcal{B}(\mathcal{F},\mwcW)$ is the set of bandlimited, finite energy signals whose spectral support is a finite union of bounded intervals,
\begin{align}
&\mathcal{B}(\mathcal{F},\mwcW)=\bigl\{x(t)\in L^2(\mathbb{R}): X(\omega)=0, \omega\notin\mathcal{F}\bigr\}\label{equ:mwc_support1}
\intertext{where}
&\mathcal{F}=\bigcup_{i=1}^{K} [a_{i},b_{i}), \quad \lvert a_{i} \rvert, \lvert b_{i} \rvert \leq \pi \mwcW . \label{equ:mwc_support2}
\end{align}

In the following description of the MWC, primes are added to the parameters to distinguish them from the parameters of the RD. 
The same letters are, however, used for similar quantities.
For example, $W/2$ denotes the bound on the harmonics of multitone signals while $\mwcW/2$ denotes the bandwidth of the multiband signals. 

Like the RD, the $i$th channel of the MWC multiplies $x(t)$ by a random signal  $p_{i}(t)$, then filters and samples the product $x(t)p_{i}(t)$ at a sub-Nyquist rate (see Figure~\ref{subfig:mwc}).
As in the original formulation, we assume each channel's filter is an ideal low pass filter, although it has been shown that the MWC can operate with non-ideal low pass filters~\cite{chen_etal2010}.
Here, we examine its original formulation to make a clearer comparison to the RD.
The signals $p_{i}(t), i=0,\dotsc,\mwcq-1$, are periodic extensions of different realisations of the continuous random process used for the RD.
Formally, let $\B{Z}=\{Z_l\}$, be a sequence of independent and identically distributed Bernoulli random variables taking values $\pm1$ with equal probability, and for some positive integer $\mwcL$, let $p(t;\B{Z})$ denote the random process
\begin{equation}
p(t;\B{Z})=Z_l, \quad t\in \Bigl[  \frac{l}{\mwcW},  \frac{l+1}{\mwcW}\Bigr),~l=0,\dotsc, \mwcL-1.
\end{equation}
Let $\B{Z}_{i}$ denotes a particular realisation of $\B{Z}$.
The signals $p_i(t)$ are then periodic extensions of the realisations $p(t;\B{Z}_{i})$ of $p(t;\B{Z})$:
\begin{equation}
p_i(t+mT_p)=p(t;\B{Z}_i),~\text{for~} t\in[0,T_p], ~m\in\mathbb{Z},
\end{equation} 
where $T_p=\tfrac{\mwcL}{\mwcW}$.
The impulse response of the ideal low pass analogue filter is $h(t)=\tfrac{\pi \mwcW}{\mwcL}\sinc(\tfrac{\pi \mwcW}{\mwcL}t)$, implying a cut-off frequency of $\tfrac{2\pi \mwcW}{\mwcL}$ radians per second.
(This is a slightly different assumption than that made in~\cite{mishali_eldar2010} where the cut-off frequency was set to $\mwcW/\mwcM$.)
Each channel samples at a rate that is $\mwcM\in\real^{+}$ times slower than the Nyquist rate, i.e. $\tfrac{1}{T_{s}}=\tfrac{\mwcW}{\mwcM}$ Hz, where $T_s$ is the channels' sampling period.
The system's average sampling rate is $\mwcq \mwcW/\mwcM$ Hz.
In~\cite{mishali_eldar2010}, Mishali and Eldar showed that a necessary condition for successful reconstruction is $\mwcM \leq \mwcL$ or that $T_s \leq T_p$.
We assume this condition holds for the MWC throughout the paper.
We also assume that the average rate is always less than the Nyquist rate ($\mwcq<\mwcM$).

\textbf{Time domain description.}
By inspection of Figure~\ref{subfig:mwc}, we obtain the following time-domain expression for a single channel of the MWC:
\begin{align}\label{equ:mwc_timedomain}
y_i(k)&=g_{i}\bigl(k \tfrac{\mwcM}{\mwcW}\bigr) \nonumber \\
&=\frac{\pi\mwcW}{\mwcL}\int_{-\infty}^{\infty}x(\tau)p_{i}(\tau)\sinc{(\tfrac{\pi \mwcW}{\mwcL}(t-\tau))}~d\tau \Biggr\rvert_{t=k \tfrac{\mwcM}{\mwcW}},
\end{align}
for all $k\in\mathbb{Z}$.
This expression corresponds to the time domain expression in~\eqref{equ:rd_timedomain} for the RD.
Like the RD, the samples $y_{i}(k)$ can be interpreted as inner products with a set of sampling functions $\{\tfrac{\pi\mwcW}{\mwcL}p_{i}(\tau)\sinc{\bigl(\tfrac{\pi\mwcM}{\mwcL}k-\tfrac{\pi\mwcW}{\mwcL}\tau\bigr)}\}$ where
\begin{equation}\label{equ:mwc_innerprod}
y_{i}(k)= \bigl\langle x(\tau), \tfrac{\pi\mwcW}{\mwcL}p_{i}(\tau)\sinc{\bigl(\tfrac{\pi\mwcM}{\mwcL}k-\tfrac{\pi\mwcW}{\mwcL}\tau\bigr)}\bigr\rangle.
\end{equation}
But in contrast to the RD, these sampling functions have finite frequency support and infinite temporal support.
In the time-frequency plane, their support partitions the space into horizontal strips of width $\mwcW/\mwcL$ Hz (see Figure~\ref{fig:time-freq_support}, right panel).
This particular set of sampling functions represents one instance of a general theory put forth by Eldar~\cite{eldar2009} to compressively sample continuous-time signals from unions of shift-invariant spaces, of which multiband signals are members.
The theory combines modern sampling theory with CS theory in such a way that samples are acquired in a typical manner by projecting the signal onto a set of sampling functions (as in~\eqref{equ:mwc_innerprod}), but CS theory is needed for reconstruction.
We do not review the details of this theory here because it does not apply to the RD.

Interpreting~\eqref{equ:mwc_timedomain} as a random filtering, we identify the time-varying impulse response as ${h_{i}(t,\tau)=\tfrac{\pi\mwcW}{\mwcL}p_{i}(\tau)\sinc{(\tfrac{\pi \mwcW}{\mwcL}(t-\tau))}}$.
Because the MWC employs an ideal low pass filter, the impulse response contains a $\sinc{ }$ function instead of a rectangular function as seen in~\eqref{equ:rd_timedomain}.
Consequently, the impulse response has infinite temporal extent in this (ideal) setting.
For the MWC, $h_{i}(t,\tau)$ is random in the same general sense as the RD's time-varying impulse response---the $\sinc{ }$ function is multiplied by a realisation of a random process.

\textbf{Frequency domain description.}
Using standard Fourier analysis techniques, Mishali and Eldar~\cite{mishali_eldar2010} derived the following frequency domain description for the $i^{\text{th}}$ channel of the MWC,
\begin{align}
&Y_{i}(e^{j\omega\frac{\mwcM}{\mwcW}})\rect{\tfrac{\mwcM}{\pi\mwcW}\omega} \nonumber \\ 
&=\frac{\mwcW}{\mwcM} \sum_{m=-\lfloor\frac{1}{2}(\mwcL+1)\rfloor+1}^{\lfloor\frac{1}{2}(\mwcL+1)\rfloor} P_{i}(m) \,X(\omega-m\tfrac{2\pi\mwcW}{\mwcL})\rect{\tfrac{\mwcL}{\pi \mwcW}\omega} \label{equ:mwc_freqdomain} \\
&=\sum_{m=-\lfloor\frac{1}{2}(\mwcL+1)\rfloor+1}^{\lfloor\frac{1}{2}(\mwcL+1)\rfloor} \beta(m)\,X(\omega-m \tfrac{2 \pi\mwcW}{\mwcL})\rect{\tfrac{\mwcL}{\pi \mwcW}\omega} \sum_{l=0}^{\mwcL-1} p_{il}\,   e^{-j\frac{2\pi}{\mwcL}ml}, \label{equ:mwc_freqdomain_a}
\end{align}
where $P_{i}(m)$ denotes the FS coefficients of $p_{i}(t)$, $\lfloor\cdot\rfloor$ denotes the floor rounding operation, $p_{il}=p_{i}(t)$ for $t\in[l/\mwcW,(l+1)/\mwcW)$ and
\begin{equation*}
\beta(m)=\begin{cases} \dfrac{\mwcW}{\mwcM} \dfrac{1-e^{-j\frac{2\pi}{\mwcL}m}}{j2\pi m}, & m\neq 0 \\
1/\mwcL, & m=0 \end{cases}.
\end{equation*}
Appendix~\ref{app:MWC} contains a slightly different derivation of~\eqref{equ:mwc_freqdomain_a} than that presented in~\cite{mishali_eldar2010}.
Comparing~\eqref{equ:mwc_freqdomain} to~\eqref{equ:rd_freqdomain}, we observe that the DTFT of the output sequences $y_{i}(k)$ can again be interpreted as the result of a random convolution, where the frequency-varying impulse response is given by $H_{i}(m,\omega)=P_{i}(m) \rect{\tfrac{\mwcL}{\pi \mwcW}\omega}$.
The spectral content of the samples $y_{i}(k)$ is expressed by the DTFT, as opposed to the DFT, because $x(t)$ is defined on the entire real line for the MWC instead of on an interval.
We also note that the scalars $\beta(m)$ are the complex conjugates of $\alpha(n)$ in~\eqref{equ:rd_linearsystem}.

\begin{figure}
\centerline{\includegraphics[width=6.5cm]{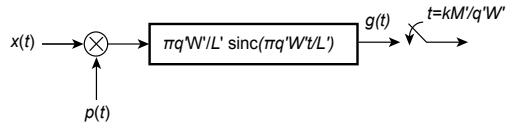}}
\caption{\small Block diagram of a single channel MWC.
To be equivalent to the multi-channel system depicted in Figure~\ref{subfig:mwc}, this system samples at a rate $\mwcq$ times faster and has a low pass filter with a cut-off frequency $\mwcq$ times greater.
Additional digital processing is also required to form the linear system in~\eqref{equ:mwc_matrixform}.}\label{fig:mwc_structure_single_channel}
\end{figure}

\textbf{Single channel MWC.}
There are two ways to collapse the MWC into an equivalent single channel system.
One can either lengthen the observation interval by a factor of $\mwcq$ (keeping all other parameters fixed), or one can consider increasing the sampling rate while maintaining the same observation interval. 
If the observation interval is lengthened, the sequence of samples from a single channel MWC can be partitioned into $\mwcq$ groups of $\mwcW T/\mwcM$, where each group of samples is thought of as the output from an individual channel in the multi-channel configuration.
Alternatively, one can set the sampling rate of a single channel MWC equal to the average rate of a multi-channel MWC, i.e., set the sampling rate to $\mwcq\mwcW/\mwcM$ Hz, and accordingly adjust the low pass filter's cut-off frequency to $\mwcq\mwcW/\mwcL$ Hz (see Figure~\ref{fig:mwc_structure_single_channel}). 
Notice that in this case we still maintain the requirement $\mwcM\leq\mwcL$.
The frequency domain description of this single channel MWC can now be obtained from~\eqref{equ:mwc_freqdomain} by substituting $\mwcM/\mwcq$ for $\mwcM$ and $\mwcL\mwcq$ for $\mwcL$:
\begin{equation}
\begin{split}
&Y(e^{j\omega\frac{\mwcM}{\mwcq\mwcW}})\rect{\tfrac{\mwcL}{\pi \mwcq\mwcW}\omega}= \\ 
&\frac{\mwcq\mwcW}{\mwcM} \sum_{m=-\lfloor\frac{1}{2}(\mwcL/\mwcq+1)\rfloor+1}^{\lfloor\frac{1}{2}(\mwcL/\mwcq+1)\rfloor} P(m) \rect{\tfrac{\mwcL}{\pi \mwcq\mwcW}\omega} \,X(\omega-m\tfrac{2\pi\mwcW\mwcq}{\mwcL}).
\end{split}
\end{equation}

\textbf{Summary.}
Equations~\eqref{equ:rd_timedomain},~\eqref{equ:rd_freqdomain},~\eqref{equ:mwc_timedomain}, and~\eqref{equ:mwc_freqdomain} all indicate that the sampling mechanisms for the RD and the MWC are based on analogue random filtering/convolution.
However, the RD's integrator and the MWC's low pass filter induce significant differences in the specific form of the random convolutions, or equivalently, in their sampling functions. 
In fact, the different filters induce bipolar time-frequency characterisations that make them well-suited for the signal models they target---an ideal integrator with a finite impulse response is well-suited to signals modeled on a finite interval and an ideal low pass filter with a finite frequency response is well-suited to signals modeled on a finite frequency band.

\section{Sparse signal models and system characteristics}\label{sect:signal_models}
This section introduces the notion of signal sparsity and briefly discusses signal reconstruction for the RD and the MWC.
We show that in practice the MWC can at best recover an approximation of the input multiband signal instead of perfectly reconstructing it.
This important fact, although not surprising, is completely missing from the current literature.
We also discuss three system characteristics related to sparsity that ultimately stem from their reliance on CS theory and algorithms. 
The first is a model sensitivity of the RD that is already a familiar limitation~\cite{tropp_etal2010,y_chi_etal_icassp2010,duarte_baraniuk2010}. 
The second concerns a possible sensitivity of the MWC with respect to the number of channels, and while again this behaviour may not be surprising, it is nevertheless important in the design of the MWC.
The third characteristic, in contrast to the second, concerns the robustness of the MWC to increases in the width of the spectral bands caused by windowing.   
The MWC results are new.

\subsection{Sparse multitone signals and RD signal reconstruction}\label{subsect:signal_models_rd}
In the original formulation of the RD, the input signal was not only modeled as a multitone signal on the observation interval, but was also assumed to be spectrally sparse~\cite{tropp_etal2010}.
A \emph{spectrally sparse multitone} signal is a multitone signal that has a small number of nonzero FS coefficients out of the $N+1$ possible (refer to~\eqref{equ:multitone_model}). 
More precisely, letting $K$ denote the number of nonzero coefficients (or equivalently the number of nonzero frequencies), a spectrally sparse multitone signal is one that satisfies $K \ll N$.

The reconstruction of a sparse multitone signal $x(t)$ from the samples $y(k)$ hinges on the matrix form of~\eqref{equ:rd_linearsystem},
\begin{equation}\label{equ:rd_matrixform}
\B{y}=\rdphi \rdpsi \,\B{X} 
\end{equation}
where $\rdphi$ is a $M\times N$ matrix of the form
\begin{equation*}
\rdphi=\begin{bmatrix}p_{0} & \ldots & p_{\frac{N}{M}-1} & 0 & & \cdots& 0 \\
0 & \cdots & 0 & \ddots & 0 & \cdots & 0\\
0 & \cdots & & 0 & p_{(M-1)\frac{N}{M}} & \ldots & p_{N-1}
\end{bmatrix}, \end{equation*}
and
\begin{align*}
\B{y}&=[y(0),\dotsc,y(M-1)]' \\
\rdpsi_{r,l}&=e^{-j\tfrac{2\pi}{N}n_{r}l} \\
\B{X}&=[\alpha(-\tfrac{N}{2}) X(-\tfrac{N}{2}),\ldots,\alpha(\tfrac{N}{2}-1) X(\tfrac{N}{2}-1)]' \\
\alpha(n_r)&=T \dfrac{e ^{j\frac{2\pi}{N}n_{r}}-1}{j2\pi n_{r}}, \quad \alpha(0)=1/W
\end{align*}
for $n_{r}=-N/2 +r$, $r=0,\dotsc, N-1$, and $l=0,\dotsc,N-1$, where the apostrophes denote transpose.

By construction,~\eqref{equ:rd_matrixform} is an underdetermined linear system of equations ($\rdphi \rdpsi$ is $M \times N$ with $M<N$; see Section~\ref{subsect:sampling_with_rd}) and underdetermined systems do not, in general, have unique solutions.
Nevertheless, CS theory has shown that with the presumed sparsity of $\B{X}$, ~\eqref{equ:rd_matrixform} can be solved by a direct application of a number of recently developed recovery algorithms, e.g., $\ell_{1}$ minimisation~\cite{candes_romberg_tao2006}, orthogonal matching pursuit~\cite{tropp2007}, or iterative hard thresholding\cite{blumensath_davies2008,blumensath_davies2009}.
In the CS literature, solving~\eqref{equ:rd_matrixform} is termed the \emph{single measurement vector} (SMV) problem.
Theoretical guarantees regarding the successful recovery of $\B{X}$ are provided in~\cite{tropp_etal2010} in terms of the degree of sparsity and the number of samples (measurements) collected.

\subsection{Sparse multiband signals and MWC signal reconstruction}\label{subsect:signal_models_mwc}
Recall that multiband signals are bandlimited, finite energy signals whose spectral support $\mathcal{F}$ is a union of bounded intervals (see~\eqref{equ:mwc_support1} and~\eqref{equ:mwc_support2}). 
A \emph{sparse} multiband signal is a multiband signal whose support has Lebesgue measure that is small relative to the overall signal bandwidth, i.e., $\lambda(\mathcal{F})\ll \mwcW$~\cite{pingfeng_phdthesis1997}.
If, for instance, all the occupied bands (intervals) have equal bandwidth $B$ Hz and the signal is composed of $K$ disjoint frequency bands, then a sparse multiband signal is one satisfying $KB\ll \mwcW$.
In the CS literature, signals having this type of ``block'' structure have been studied in various settings; the central question being whether this additional signal structure reduces the minimum number of samples required to reconstruct the original signal (see e.g.,~\cite{stojnic_etal2009,baraniuk_etal2010,eldar_etal2010}). 

MWC signal reconstruction centres on the matrix form of~\eqref{equ:mwc_freqdomain},
\begin{equation}\label{equ:mwc_matrixform}
\B{Y}(\omega)=\B{\Phi}\B{\Psi}\B{S}(\omega)
\end{equation}
where
\begin{align*}
\B{Y}(\omega)&=[Y_{0}(\omega),\dotsc,Y_{\mwcq-1}(\omega)]' \\
Y_{i}(\omega)&=Y_{i}(e^{j\omega\frac{\mwcM}{\mwcW}}) \rect{\tfrac{\mwcM}{\pi\mwcW}\omega} \\ 
\B{\Phi}_{i,l}&=p_{il} \\
\B{\Psi}_{l,r}&= e^{-j\tfrac{2\pi}{\mwcL}lm_{r}} \\
\B{S}(\omega)&=[S_{0}(\omega),\dotsc,S_{\mwcL-1}(\omega)]' \\
S_{r}(\omega)&= \beta(m_r)\, \rect{\tfrac{\mwcL}{\pi \mwcW}\omega}\, X(\omega-m_{r}\tfrac{2\pi\mwcW}{\mwcL}) \\
\beta(m_r)&= \frac{\mwcW}{\mwcM} \dfrac{1-e^{-j\frac{2\pi}{\mwcL}m_{r}}}{j2\pi m_{r}}, \quad \beta(0)=1/\mwcL
\end{align*}
for $i=0,\dotsc,\mwcq-1$, $l=0,\dotsc,\mwcL-1$, $r=0,\dotsc,\mwcL-1$ and $m_{r}=-\lfloor\frac{1}{2}(\mwcL+1)\rfloor+1+r$.
Like~\eqref{equ:rd_matrixform}, this linear system of equations is underdetermined since we assume $\mwcq<\mwcM \leq \mwcL$ (see Section~\ref{subsect:mwc}).
The vector $\B{S}(\omega)$ is sparse in the sense that most of its elements (segments of $X(\omega)$) do not contain the occupied frequency bands that comprise $x(t)$.
Equation~\eqref{equ:mwc_matrixform} can also be derived from the single channel MWC, although one has to first extract $\mwcq$ lower rate sample sequences from the single higher rate output sequence.
We refer the reader to~\cite{mishali_eldar2010} for details regarding the extra processing steps.

In practice the linear system in~\eqref{equ:mwc_matrixform} cannot in general be computed because it theoretically requires an infinite amount of data.
To see the point, consider the inverse DTFT of~\eqref{equ:mwc_matrixform}.
It immediately follows that the inverse DTFTs of the spectra $Y_{i}(\omega)=Y_{i}(e^{j\omega\frac{\mwcM}{\mwcW}})\rect{\tfrac{\mwcM}{\pi\mwcW}\omega}$ are the time domain sequences $\{y_{i}(k),k\in\mathbb{Z}\}_{i}$.
If one interprets the spectral segments $S_{r}(\omega)=\beta_{r} \rect{\tfrac{\mwcL}{\pi \mwcW}\omega}\, X(\omega-m_{r}\tfrac{2\pi\mwcW}{\mwcL})$ on the right hand side of~\eqref{equ:mwc_matrixform} as single periods of periodic spectra, it follows that their inverse DTFT are the sequences $\{\gamma_{r}(k),k\in\mathbb{Z}\}_{r}$ where
\begin{equation}
\gamma_{r}(k)=\frac{\mwcL}{2\pi\mwcW} \int_{-\pi \mwcW/\mwcL}^{\pi \mwcW/\mwcL} S_{r}(\omega) e^{j\frac{\mwcL}{\mwcW}k\omega}~d\omega,
\end{equation}
and
\begin{equation}\label{equ:spectal_slice_DTFT}
S_{r}(\omega)=\sum_{k=-\infty}^{\infty} \gamma_{r}(k) e^{-j\tfrac{\mwcL}{\mwcW}k\omega} , \quad \omega\in[-\tfrac{\pi\mwcW}{\mwcL}, \tfrac{\pi\mwcW}{\mwcL}].
\end{equation}
The DTFT transform pair of~\eqref{equ:mwc_matrixform} is therefore the linear system,
\begin{equation}\label{equ:mwc_imv}
\B{Y}=\mwcphi\mwcpsi \B{S}
\end{equation}
where $\mwcphi$ and $\mwcpsi$ are as in~\eqref{equ:mwc_matrixform} but $\B{Y}$ and $\B{S}$ are now \emph{infinite} column matrices: the rows of $\B{Y}$ are the sequences $y_{i}(k), k\in\mathbb{Z}$, and the rows of $\B{S}$ are the sequences $\gamma_{r}(k), k\in\mathbb{Z}$.
The matrix $\B{S}$ is described as being \emph{jointly sparse} because most of its rows are zero since the zero-valued elements of $\B{S}(\omega)$ correspond to zero-valued sequences $\gamma_{r}(k)$ (rows of $\B{S}$).
In general, a matrix $\B{Z}$ is said to be $K$ joint sparse if there are at most $K$ rows in $\B{Z}$ that contain nonzero elements.
The recovery of $\B{S}$ from the measurements $\B{Y}$ in~\eqref{equ:mwc_imv} is called an \emph{infinite measurement vector} (IMV) problem~\cite{mishali_eldar2008,eldar2009,mishali_eldar2010} because the columns of $\B{Y}$ are viewed as CS measurements (via the measurement matrix $\mwcphi\mwcpsi$) of a collection of vectors that share a common sparse support.

The limitation of only ever collecting a finite number of samples truncates the rows of $\B{Y}$ and causes the IMV problem in~\eqref{equ:mwc_imv} to become a so-called \emph{multiple measurement vector} (MMV) problem~\cite{cotter_etal2005,greedy_pursuit_tropp_etal2006,convex_relax_tropp_etal2006,chen_huo2006,mishali_eldar2008,davies_eldar2010}, where the goal is to recover a finite number of the columns of the jointly sparse matrix $\B{S}$ corresponding to the finite number of acquired samples.
Using existing CS methods, this MMV problem can be solved exactly, or with exceedingly high probability, provided the matrix $\mwcphi\mwcpsi$ satisfies certain conditions and that enough samples are collected relative to the joint sparsity of $\B{S}$.
The solution, however, is in general a linear approximation to the true spectral slices $Y_{i}(\omega)$ because the solution only recovers a finite number of coefficients $\gamma_{r}(k)$ in~\eqref{equ:spectal_slice_DTFT} (see~\cite{mallat1999} for information about linear approximations). 
This fact is in contrast to the sparse multitone signal model that is parameterised by a finite number of parameters and thus only requires a finite number of samples for perfect signal reconstruction.

In~\cite{mishali_eldar2010} and~\cite{mishali_eldar2008}, Mishali and Eldar proposed a two step reconstruction process termed the ``continuous-to-finite block'' that provably recovers $x(t)$ exactly given an infinite amount of data, or in other words, recovers $x(t)$ to an arbitrary precision given sufficient data.
The first step recovers the joint support of $\B{S}$ by solving an associated MMV problem, and the second step uses the recovered support to reduce the dimension of the measurement matrix $\mwcphi\mwcpsi$ such that a unique least squares solution can be found.
We stress that even if this two step process perfectly solves the MMV problem derived from~\eqref{equ:mwc_imv}, the solution can only, in general, approximate $x(t)$ for a finite number of samples.

\subsection{RD sensitivity to basis mismatch}
The ability of CS recovery algorithms to recover the FS coefficients in~\eqref{equ:rd_matrixform} depends fundamentally on the sparsity of $\B{X}$, or equivalently, on whether $x(t)$ has a sparse FS representation in the observation window $[0,T]$ and on the number of acquired measurements.
For the RD, the sparsity level of $\B{X}$ (number of nonzero entries) in~\eqref{equ:rd_matrixform} not only changes with the number of tones comprising $x(t)$, but can also be increased if there is a mismatch between the Fourier basis in which $x(t)$ is actually sparse and the basis in which $x(t)$ is modeled.
In fact it is known that $x(t)$ may have a sparse Fourier expansion using $\{\exp{(j\tfrac{2\pi}{T}nt)}\}_{n}$ but may not have a sparse expansion using $\{\exp{(j\tfrac{2\pi}{T+\delta}nt)}\}_{n}$, where $\delta\in\real$ is some small perturbation~\cite[p.379-380]{mallat1999}.
The implication for the RD is that in a blind sensing scenario, where the frequencies of the tones are not known, model mismatches are likely and if the sparsity level of $\B{X}$ rises above a required level, CS recovery may be jeopardized.
This possibility has been described as a sensitivity of the RD because a small basis mismatch (small $\delta$) can lead to significant reconstruction errors~\cite{y_chi_etal_icassp2010}.
This sensitivity was acknowledged by Tropp et al. in~\cite{tropp_etal2010}, highlighted in~\cite{mishali_eldar_dcc2010,mishali_eldar_elron2011} and studied in~\cite{y_chi_etal_icassp2010,duarte_baraniuk2010}.
Recent results, however, could potentially alleviate the problem.
In~\cite{candes_etal2011}, Cand\'{e}s et al. showed that sparse multitone signals, defined in terms of an oversampled DFT dictionary, can be effectively recovered from undersampled data. 
Because this model can sparsely represent multitone signals on $[0,T]$ as well as on $[0,T+\delta]$, these results suggest that a modified RD could be robust to basis mismatch. 
Similarly, Duarte and Baraniuk~\cite{duarte_baraniuk2010} proposed a heuristic solution that marries model-based CS~\cite{baraniuk_etal2010}, redundant DFT frames, and standard spectral estimation techniques.

\subsection{Potential MWC sensitivity to the number of channels}
\begin{figure}
\begin{minipage}{.48\textwidth}
\centerline{\includegraphics[width=6cm]{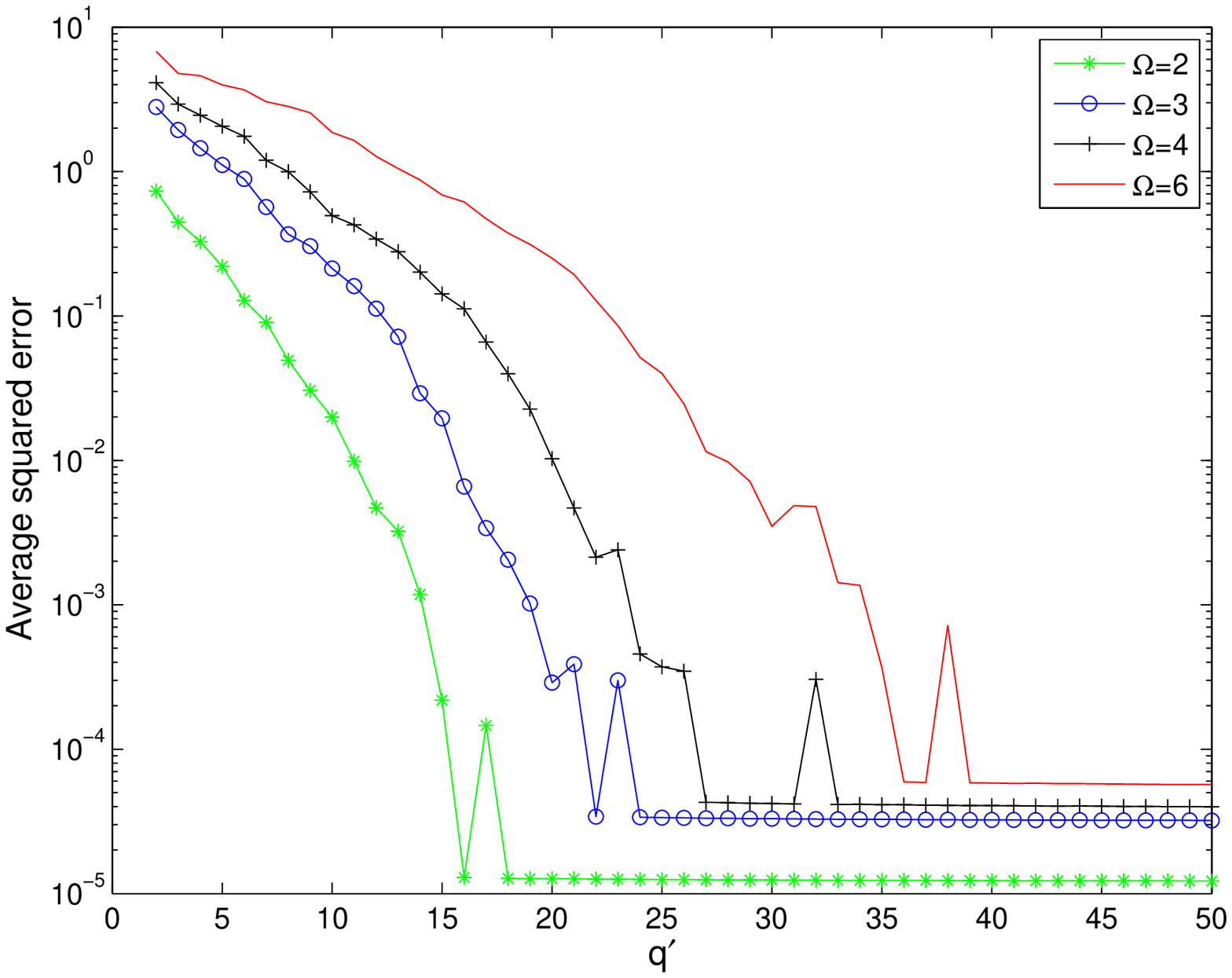}}
\end{minipage}
\begin{minipage}{.48\textwidth}
\centerline{\includegraphics[width=6cm]{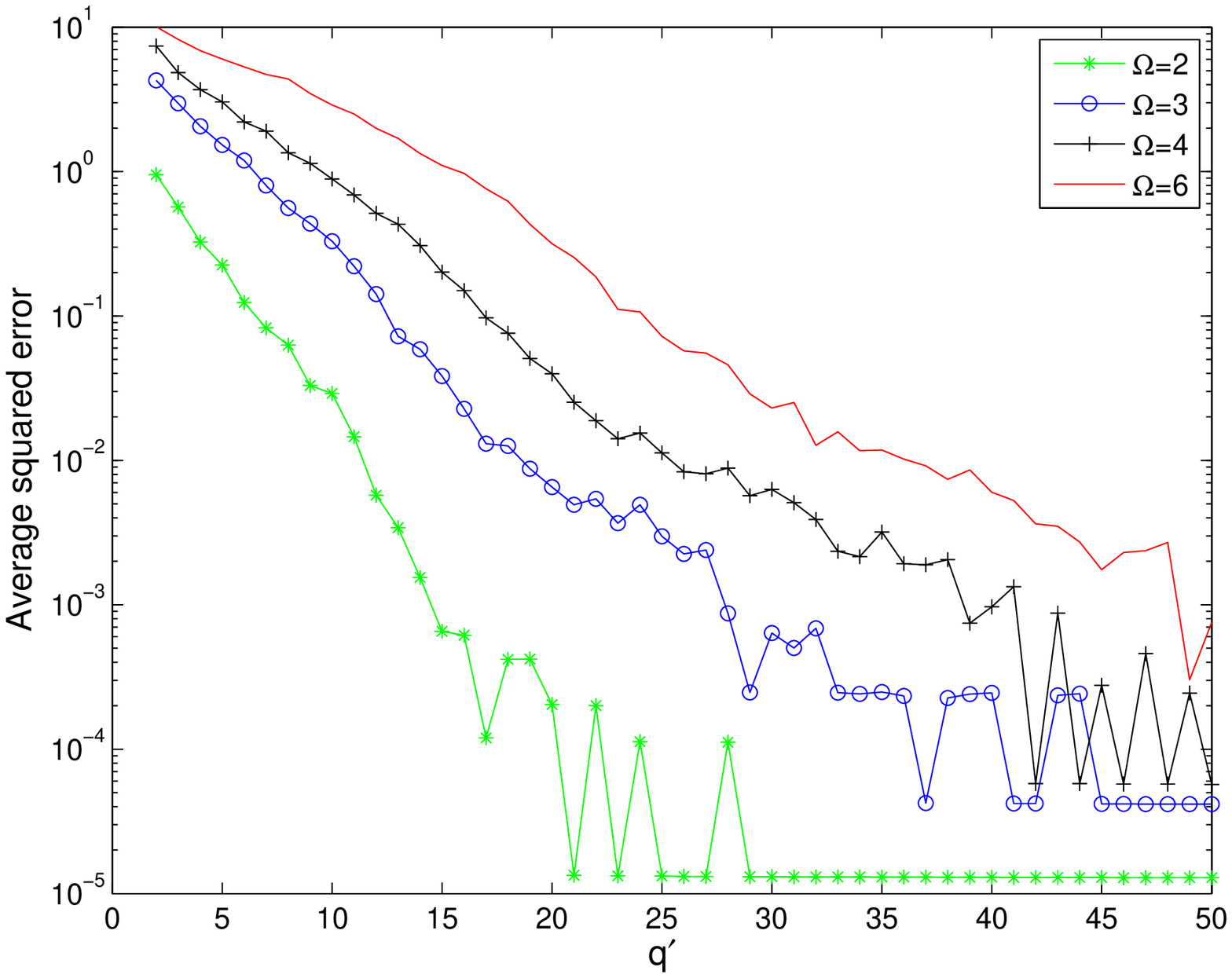}}
\end{minipage}
\caption{\small Reconstruction error as a function of the number of channels $\mwcq$ for the MWC. $\Omega$ denotes the joint sparsity $\lvert\text{supp}\,\B{S}\rvert$ of $\B{S}$ and S-OMP is used to recover the input signal's support with $T=4$ sec, $\mwcW=500$ Hz, $\mwcL=50$, $\mwcM=20$. The plots show the rapid increase in error as $\mwcq$ falls below a critical point. For the cases shown, the error increases \emph{exponentially} as $\mwcq$ decreases. Left: active bands have equal amplitude; Right: unequal amplitudes.}\label{fig:mwc_q_exp_error}
\end{figure}

\begin{figure}
\begin{minipage}{.48\textwidth}
\centerline{\includegraphics[width=6.5cm]{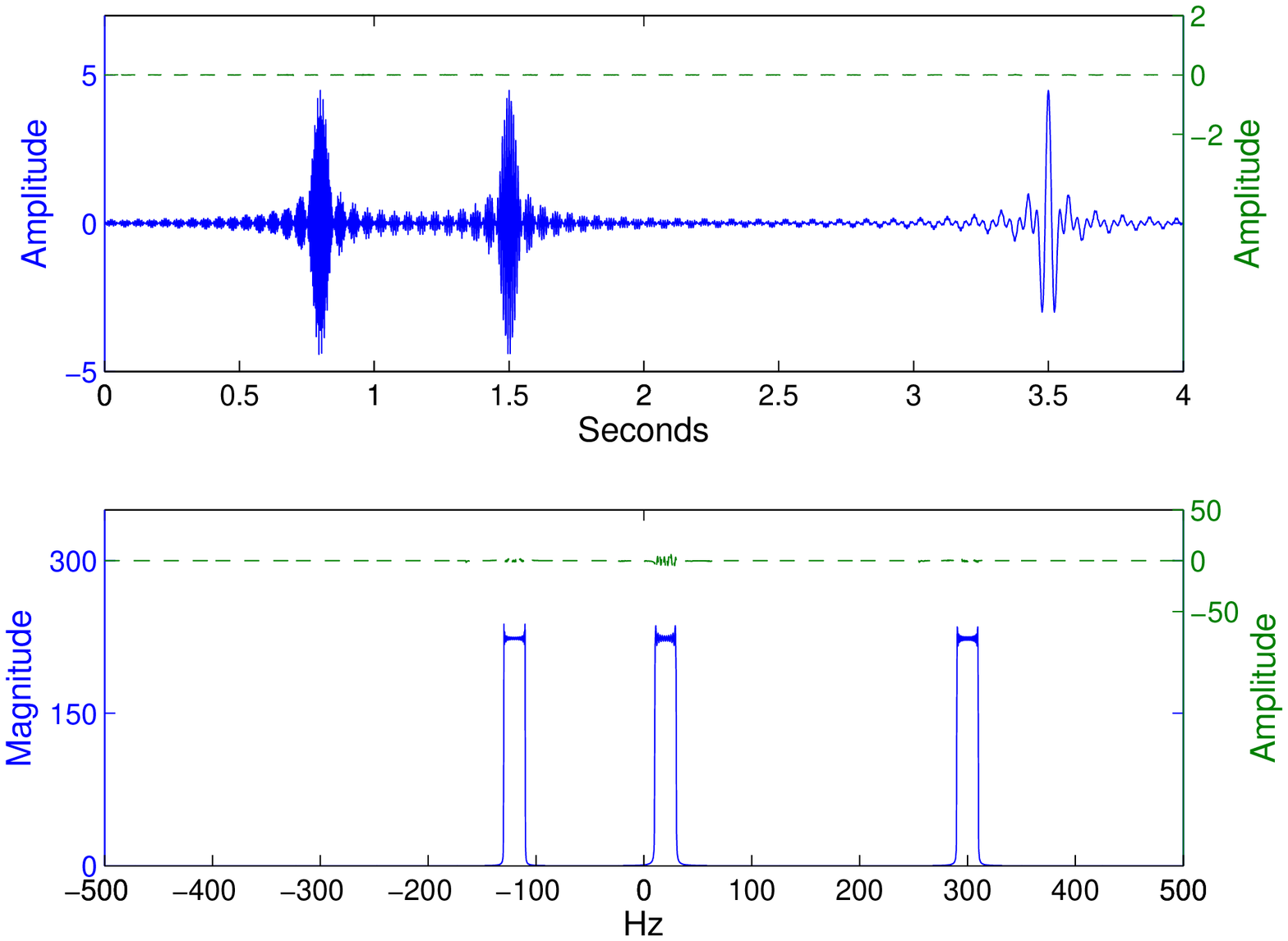}}
\end{minipage}
\begin{minipage}{.48\textwidth}
\centerline{\includegraphics[width=6.5cm]{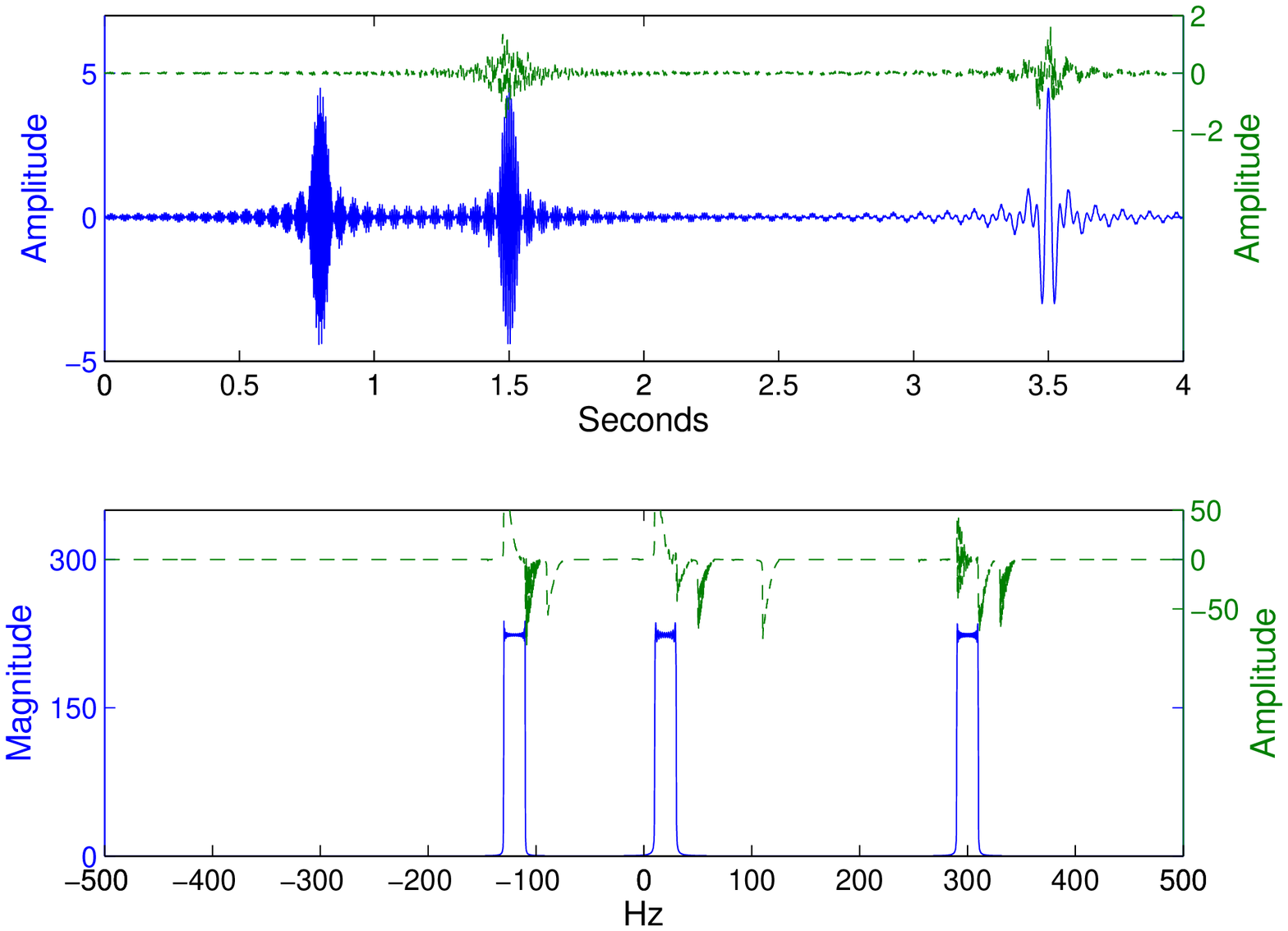}}
\end{minipage}
\caption{\small Results of two specific cases from the experiment in Figure~\ref{fig:mwc_q_exp_error} ($\Omega=3$, equal amplitudes). The time and frequency domains of the input multiband signals are shown along with the difference signal between the input and reconstructed signals (green raised plots). Left: $\mwcq=25$, squared error$=3\times10^{-5}$; Right: $\mwcq=18$, squared error$=0.001$.} \label{fig:mwc_q_exp_timefreq}
\end{figure}

Because of its reliance on CS, the MWC reconstruction algorithm inherits the CS conditions of successful reconstruction.
One such condition relates the number of CS measurements per unit time (number of rows of $\B{Y}$ in~\eqref{equ:mwc_imv}) to the support of $\B{S}$.
Assuming that the matrix $\mwcphi\mwcpsi$ has maximal rank $\mwcq$, a necessary condition for unique recovery of $\B{S}$ from the measurements is $\mwcq>2 \lvert\text{supp}\,\B{S}\rvert$, where $\lvert\text{supp}\,\B{S}\rvert$ denotes joint sparsity of $\B{S}$~\cite{mishali_eldar2010,chen_huo2006}.
Knowing that $\mwcq$ also equals the number of channels in the MWC, it is reasonable to want to minimise $\mwcq$ so as to minimise the processing and hardware complexity of the MWC, i.e. set it as close to $2\lvert \text{supp}\,\B{S}\rvert$ as possible while still ensuring successful recovery.
(In practice, this lower bound changes depending on the specific CS algorithm employed.)
With this in mind, it is natural to ask how MWC's performance behaves around this condition boundary.
Clearly, beyond the theoretical limit of $2\lvert \text{supp}\,\B{S}\rvert$, it is impossible to guarantee recovery of the signal's support.
Consequently, one expects the performance to degrade beyond this point. 
Figures~\ref{fig:mwc_q_exp_error} and~\ref{fig:mwc_q_exp_timefreq} show that performance (as measured as squared error) can indeed decay rapidly, in fact \emph{expontially fast}, as a function of the number of channels $\mwcq$.
In the example, we consider a sparse multiband signal bandlimited to $500$ Hz that is sampled by a MWC with a spectral resolution of $20$ Hz ($\mwcL=50$) and a channel sampling rate of $50$ Hz ($\mwcM=20$) for various sparsity levels and values of $\mwcq$.
The average squared error is computed as $\tfrac{1}{T\mwcW} \norm{\B{x}-\widehat{\B{x}}}_{2}^{2}$, where $\B{x}$ is a digitally simulated sparse multiband signal, $\widehat{\B{x}}$ is the reconstructed signal, and $\norm{\cdot}_{2}^{2}$ denotes the $\ell_{2}$ norm.
Figure~\ref{fig:mwc_q_exp_error} shows the results when the simultaneously orthogonal matching pursuit algorithm (S-OMP) is used to recover the signal's support.
For each value of $\mwcq$, we report an error that is averaged over $100$ trials with each trial using a different randomly generated $\mwcphi$ matrix.
Note that when each band has equal amplitude (left panel) degradation begins at roughly $\mwcq=2 \lvert \text{supp}\,\B{S}\rvert \log{(\mwcL)}$, which is consistent with S-OMP~\cite{greedy_pursuit_tropp_etal2006}.
When the band amplitudes are randomly chosen (right panel) the turning points are roughly proportional to $2 \lvert \text{supp}\,\B{S}\rvert \log{(\mwcL)}$.
Figure~\ref{fig:mwc_q_exp_timefreq} shows the original and reconstructed signals for one specific case. 

The point of this example is to show that when the number of channels is near its theoretical limit, a small change in its value can lead to dramatic performance decreases for the MWC.
The exact degree of degradation is problem dependent however.


\subsection{MWC robustness to windowing}
\begin{figure}
\centerline{\includegraphics[width=6cm]{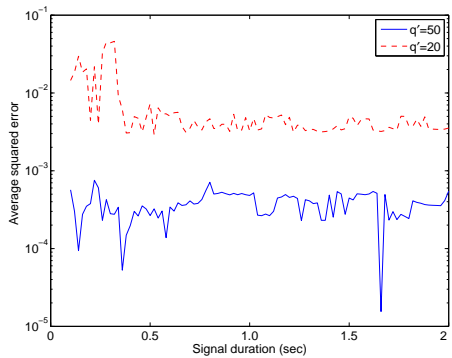}}
\caption{\small Reconstruction error as a function of signal duration for the MWC. A four band multiband signal bandlimited to $500$ Hz was systematically windowed and sampled by a MWC ($\mwcq=20,50$, $\mwcL=50, \mwcM=20$) over an observation interval of 2 sec. Each point represents a squared error averaged over 50 trials. Note that the error remains relatively constant over a wide range of signal durations. }\label{fig:mwc_wind_sen_exp_error}
\end{figure}

\begin{figure}
\begin{minipage}{.48\textwidth}
\centerline{\includegraphics[width=6.5cm]{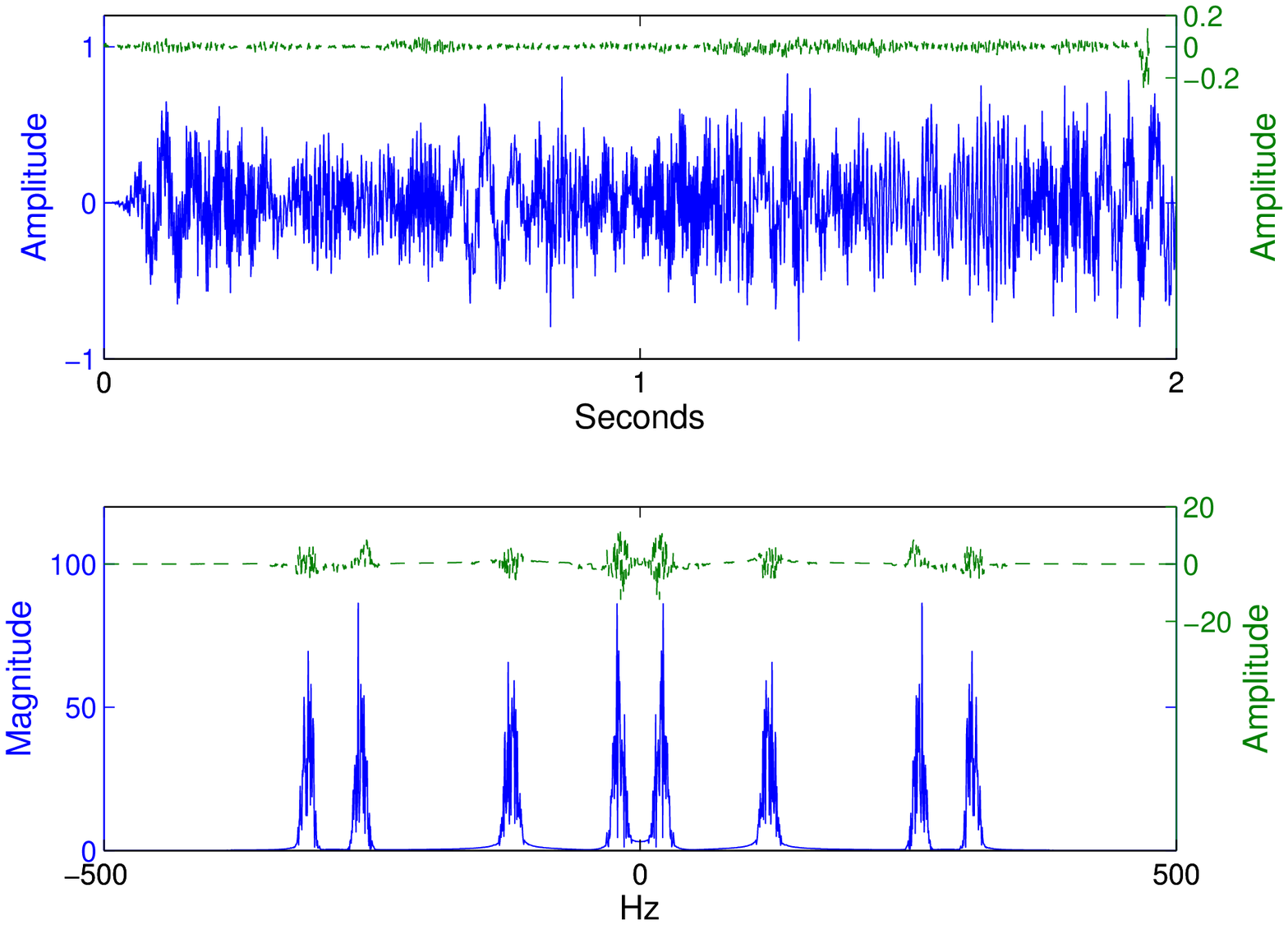}}
\end{minipage}
\begin{minipage}{.48\textwidth}
\centerline{\includegraphics[width=6.5cm]{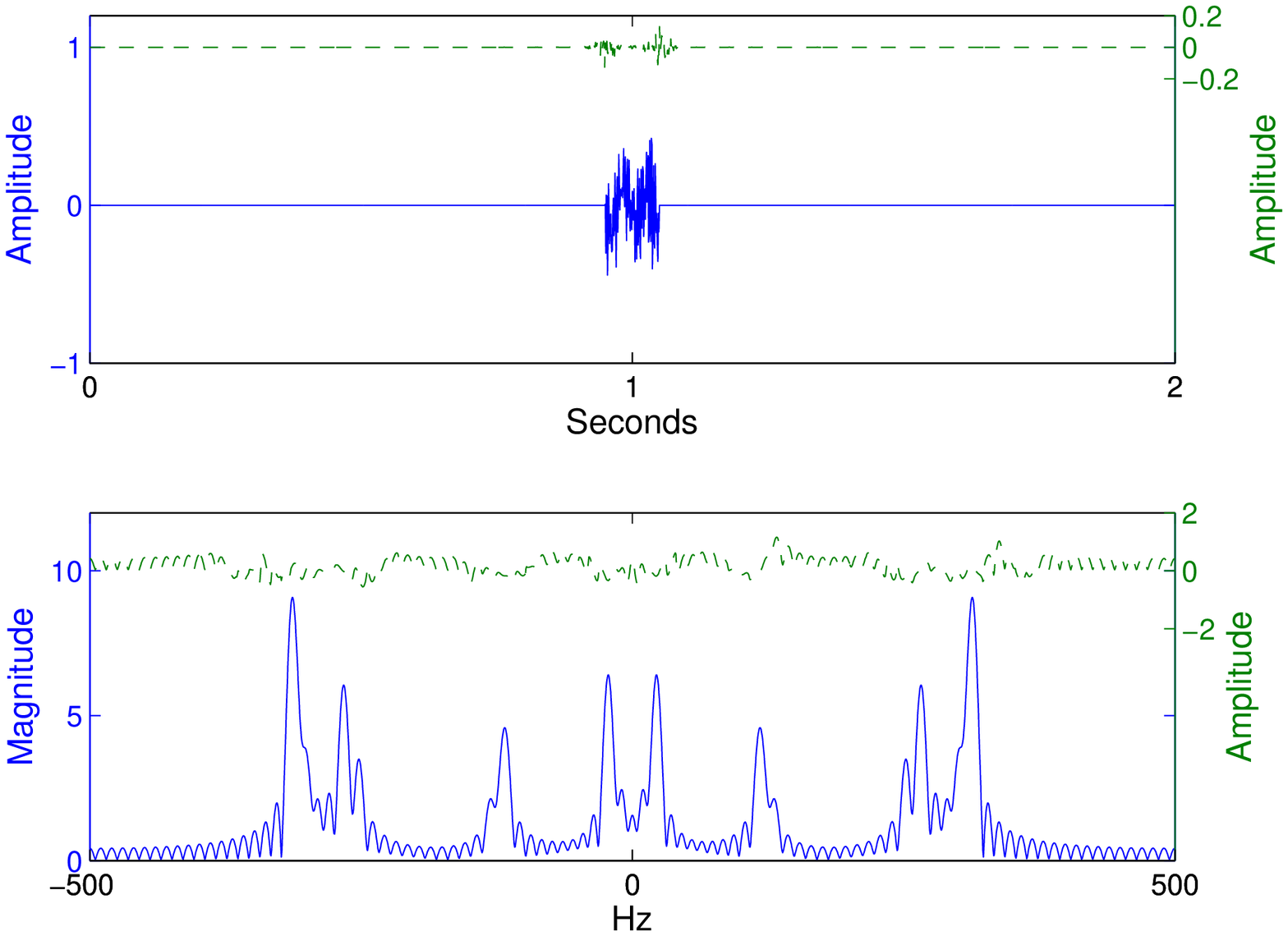}}
\end{minipage}
\caption{\small Results of two specific cases from the MWC experiment in Figure~\ref{fig:mwc_wind_sen_exp_error} ($\mwcq=50$). The time and frequency plots of two windowed multiband signals are show. They represent the two extreme cases: full duration (2 sec), short duration (0.1 sec). The raised plots (green) are the difference signals between the original and reconstructed signals. Note that the spread of the spectrum has little effect on reconstruction.} \label{fig:mwc_wind_sen_exp_timefreq}
\end{figure}

Let $x(t)$ be a sparse multiband signal with FT $X(\omega)$ and let $z(t)$ be a windowed version of $x(t)$,
\begin{equation*}
z(t)=x(t)w(t),
\end{equation*}
where $w(t)$ is an indicator function of some sub-interval of the observation interval.
Consider the situation where $z(t)$ is the input signal to the MWC, but where the MWC is designed for a multiband signal, i.e., as described in Section~\ref{subsect:mwc}.
Because time limited signals cannot be bandlimited, this case represents a model mismatch scenario where essentially the input signal has shorter duration than expected. 
The spectrum of $z(t)$ equals $X(\omega)$ convolved with a sinc function, thus the windowing spreads the original spectrum $X(\omega)$~\cite{roberts_mullisDSP}.
One might therefore expect that if the spectrum were sufficiently spread, the sparsity of $\B{S}(\omega)$ would increase to a point where the condition $\mwcq>2 \lvert\text{supp}\,\B{S}\rvert$ is violated and performance would degrade rapidly as seen above in Figure~\ref{fig:mwc_q_exp_error}.
However, experimental results show that this is not necessarily the case.
Figure~\ref{fig:mwc_wind_sen_exp_error} shows that the MWC can be robust to a wide range of signal durations.
In fact, for the two values of $\mwcq$ shown, the error remains relatively constant for signal durations ranging from $100\%$ to $5\%$ of the observation interval. 
Observe that for $\mwcq=20$, reconstruction does eventually break down (signal duration less than $0.4$ sec), but these signals represent, in some sense, those that maximally mismatch the multiband model because of their small compact support.
Figure~\ref{fig:mwc_wind_sen_exp_timefreq} shows time and frequency plots, along with the reconstruction errors, for two specific examples for long and short duration signals.

\section{Sampling continuous-time block-sparse signals}\label{sect:more_models}
This section ends with a description of a novel sampling system whose inspiration comes from a simple exercise: we exchange the signal models for the RD and the MWC and analyze the relationship between the input and output signals.
In particular, we consider sampling sparse multitone signals with the MWC and sparse multiband signals with the RD. 
We discover that the MWC uses what we call ``block-convolution'' to successfully sample and recover block sparse signals.
We also discover that the RD does not use block convolution and thus cannot, without difficulty, sample multiband signals.
In Section~\ref{subsect:timeblocksparse}, we apply this concept in a new way and propose a system that can sample and recover continuous-time block-sparse signals at low sampling rates.

\subsection{MWC with sparse multitone inputs}\label{subsect:MWCmultitone}
Consider the problem of using a MWC to sample and recover a sparse multitone signal instead of sparse multiband signal.
Let $x(t)$ be a sparse multitone signal on the observation interval $[0,T]$ with bounded harmonics $-\tfrac{W}{2}\leq \frac{n}{T} < \tfrac{W}{2}$ and denote by $X(n)$ the FS coefficients of $x(t)$ on $[0,T]$ (cf.~\eqref{equ:multitone_model}).
Let the random waveforms $p_{i}(t), i=0,\dotsc,\mwcq-1,$ be as defined in Section~\ref{subsect:mwc}, but with $\mwcW$ replaced by $W$; their common period $T_p$ thus equaling $\tfrac{\mwcL}{W}, \mwcL\in \mathbb{Z}^{+}$. 
Let the ideal low pass filter have impulse response $h(t)=\tfrac{\pi W}{\mwcL}\sinc(\tfrac{\pi W}{\mwcL}t)$ and let the sampling period be $T_s=\tfrac{\mwcM}{W}$.
For this problem and for this section only, we assume that the duration of the observation interval is an integer multiple of $T_p$, i.e., $T=N\tfrac{\mwcL}{W}, N\in \mathbb{Z}^{+}$.
We also assume, for ease of exposition, that the period of $p_{i}(t)$ equals the sampling period ($T_p=T_s$ or that $\mwcL=\mwcM$).
These assumptions are not necessary, but help make this section's message clear and easier to understand.
Lastly, to make the problem meaningful, we assume $T\geq T_s$.

In Appendix~\ref{app:MWCmultitone}, we derive the following expression relating the FS coefficients of $x(t)$ to the DFT coefficients of the output samples,
\begin{equation}
Y_{i}(n)=\sum_{m=-\lfloor\frac{1}{2}(\mwcL+1)\rfloor+1}^{\lfloor\frac{1}{2}(\mwcL+1)\rfloor} \eta(m) \,X(n-Nm)\rect{\tfrac{2}{N}n} \sum_{l=0}^{\mwcL-1} p_{il} \, e^{-j\frac{2\pi}{\mwcL}lm}, \label{equ:mwc_tone_b}
\end{equation}
where $p_{il}=p_{i}(t)$ for $t\in[l/W,(l+1)/W)$ and
\begin{equation*}
\eta(m)=\begin{cases}
\dfrac{1}{N}\dfrac{1-e^{-j\frac{2\pi}{\mwcL}m}}{j2\pi m} & m\neq 0\\
1/\mwcL & m=0
\end{cases}
\end{equation*}
for $-N/2\leq n < N/2$ and $i=0,\dotsc,\mwcq-1$.
This expression is analogous to the frequency domain description of the RD given by~\eqref{equ:rd_freqdomain} and is what~\eqref{equ:mwc_freqdomain} becomes assuming a sparse multitone signal model and $\mwcL=\mwcM$. 
In matrix form,~\eqref{equ:mwc_tone_b} becomes the MMV problem,
\begin{equation}\label{equ:mwc_multitone_mmv}
\B{Y}=\mwcphi \mwcpsi \B{S},
\end{equation}
where 
\begin{align*}
\B{Y}_{i,v}&=Y_{i}(n_{v}) \\
\B{\Phi}_{i,l}&=p_{il} \\
\B{\Psi}_{l,r}&= e^{-j\tfrac{2\pi}{\mwcL}lm_{r}} \\
\B{S}_{r,v}&= \eta(m_r) X(n_{v}-Nm_{r})\rect{\tfrac{2}{N}n_{v}} \\
\eta(m_r)&= \frac{1}{N}\frac{1-e^{-j\frac{2\pi}{\mwcL}m_{r}}}{j2\pi m_{r}}, \quad \eta(0)=1/\mwcL,
\end{align*}
for $i=0,\dotsc,\mwcq-1$, $l=0,\dotsc,\mwcL-1$, $v=0,\dotsc,N-1$,  $n_{v}=-\bigl \lfloor \tfrac{N}{2}\bigr \rfloor +v$, $r=0,\dotsc,\mwcL-1$ and $m_{r}=-\big\lfloor\frac{1}{2}(\mwcL+1)\big\rfloor+1+r$.
In contrast to sampling a sparse multiband signal, this MWC MMV problem does not result from truncation, rather its finiteness derives from the fact that multitone signals are finitely parameterised.
The Fourier components of $x(t)$ can be recovered by solving~\eqref{equ:mwc_multitone_mmv} using several existing CS algorithms including greedy algorithms~\cite{greedy_pursuit_tropp_etal2006}, mixed norm approaches~\cite{convex_relax_tropp_etal2006}, MUSIC based recovery algorithms and, in particular, the approach proposed by Mishali and Eldar in~\cite{mishali_eldar2010} and~\cite{mishali_eldar2008}.
One can thus successfully sample and recover sparse multitone signals with the MWC.
This fact is somewhat surprising since the MWC is designed to sample multiband signals, not multitone signals.

More importantly, this example highlights a property of the MWC that differentiates it from the RD and allows it to successfully sample and recover multitone signals, as well as multiband signals.
Specifically, the convolution in~\eqref{equ:mwc_tone_b} involves shifts of $X(n)$ by integer multiplies of $N$ that, when greater than one, yields a ``block convolution''.
By block-convolution, we mean every DFT coefficient $Y_{i}(n)$ in~\eqref{equ:mwc_tone_b} is a linear combination of finite segments of $X(n)$.
Block-convolution is also seen in~\eqref{equ:mwc_freqdomain} where the FT of a multiband signal is shifted by integer multiplies of $\tfrac{\mwcW}{\mwcL}$.
In contrast, it is not seen in~\eqref{equ:rd_freqdomain} where the frequency shifts describing the RD are by one.
This aspect of the MWC allows the construction of a linear system like~\eqref{equ:mwc_matrixform} and~\eqref{equ:mwc_multitone_mmv} that describe the original spectrum in terms of linear combinations of these blocks. 
The blocks themselves represent a partition of the frequency axis that effectively discretises a signal's spectrum.
In Section~\ref{subsect:timeblocksparse} below, we incorporate this property into a multi-channel random convolution system that samples and approximately recovers continuous-time block-sparse signals, the time domain analogue of sparse multiband signals.

When the observation interval, the period of $p_{i}(t)$, and the sampling period are equal, i.e. when $N=1$ and $\mwcL=\mwcM$,~\eqref{equ:mwc_multitone_mmv} collapses to a SMV problem, where the matrices $\B{Y}$ and $\B{S}$ become the vectors $[Y_{0}(0),\dotsc, Y_{\mwcq-1}(0)]^{'}$ and $[\eta(m_0) X(-m_{0}),\dotsc,\eta(m_{\mwcL-1})X(-m_{\mwcL-1})]^{'}$ respectively.
Note that in this special case, the MWC and the RD produce equivalent SMV problems, the only difference being the timing in how the samples are acquired---the MWC collects a measurement vector in parallel (each channel samples once) while the RD collects its samples sequentially.

\subsection{RD with sparse multiband inputs}\label{subsect:rd_with_multiband}
When a multiband signal is the assumed signal model for the RD, the system fails to produce a single measurement vector problem whose solution recovers $x(t)$.
To be more concrete, let $x(t)$ be a sparse multiband signal bandlimited to $\mwcW/2$ Hz with a fixed support $\mathcal{F}$.
Consider a RD parameterised by $M$ that samples $x(t)$ on $[0,T]$ and let $p(t)$ be as described in Section~\ref{subsect:sampling_with_rd}.
A similar analysis to that contained in Appendix~\ref{app:RD} leads to the expression
\begin{equation*}
\begin{split}
y(k)=\frac{1}{2\pi}\sum_{m=0}^{N'/M-1}& p_{k\tfrac{N'}{M}+m} \\
& \int_{-\pi \mwcW}^{\pi \mwcW} X(\omega)\, \frac{e^{j\frac{\omega}{\mwcW}}-1}{j\omega}\, e^{j\omega\bigl(k\tfrac{N'}{M}+m\bigr)}~d\omega,
\end{split}
\end{equation*}
where here $N'$ represents the number of Nyquist periods within the observation window ($N'=T\mwcW$).
Here, for simplicity, we assume $N'$ is a positive integer.
This expression relates the time domain output samples $y(k)$ to the Fourier spectrum of $x(t)$.
To construct a finite dimensional linear system consistent with the RD formulation, one could approximate the integral by discretizing $\omega$: 
\begin{equation*}
\begin{split}
y(k)\approx \frac{1}{2\pi} \sum_{m=0}^{N'/M-1} & p_{k\tfrac{N'}{M}+m} \\
&\sum_{i=0}^{D-1} X(\omega_i) \frac{e^{j\frac{\omega_i}{\mwcW}}-1}{j\omega_i}\, e^{j\omega_{i}\bigl(k\tfrac{N'}{M}+m\bigr)} ~\delta_{\omega},
\end{split}
\end{equation*}
where $\delta_{\omega}=\tfrac{2\pi W}{D}$ for some positive integer $D$ and $\omega_{i}=-\pi W+\delta_{\omega}(i+1/2)$.
One can then see that in comparison to~\eqref{equ:rd_linearsystem} (and also~\eqref{equ:rd_a} in Appendix~\ref{app:RD}), this expression describes a SMV problem whose dimensions grow linearly with $D$.
Clearly, in cases where one wants to closely approximate the integral or finely discretize the $\omega$ axis, the size of the SMV problem could become computationally unwieldy.
In fact, examples from~\cite{mishali_eldar_elron2011} show that naively modeling a multiband signal as a multitone signal can lead to computationally prohibitive or even intractible CS problems given current technology.
In comparison to the MWC, one reason for this difficulty is that the RD \emph{does not use block-convolution}.

\subsection{Continuous-time block-sparse sampler}\label{subsect:timeblocksparse}
The class of \emph{continuous-time block signals} $\mathcal{G}(\mathcal{T},t_{0})$ is the set of continuous-time, real-valued, finite energy signals whose support is a finite union of bounded intervals,
\begin{equation*}
\mathcal{G}(\mathcal{T},t_{0})=\bigl\{x(t)\in L^{2}([0,t_{0}])\cap C([0,t_{0}]): x(t)=0, t\notin \mathcal{T} \bigr\} \label{def:time_block_signals}
\end{equation*}
where $C([0,t_{0}])$ denotes the set of continuous functions on $[0,t_0]$ and
\begin{equation*}
\mathcal{T}=\bigcup_{i=1}^{K} [a_{i},b_{i}), \quad 0\leq a_{i}, b_{i} \leq t_{0}<\infty.
\end{equation*}
A \emph{continuous-time block-sparse signal} is a continuous-time block signal whose support has Lebesgue measure that is small relative to the signal's overall duration, i.e., $\lambda(\mathcal{T}) \ll t_{0}$.

\begin{figure}
\subfigure[ ]{\begin{minipage}{.48\textwidth}
\centerline{\includegraphics[width=4.5cm]{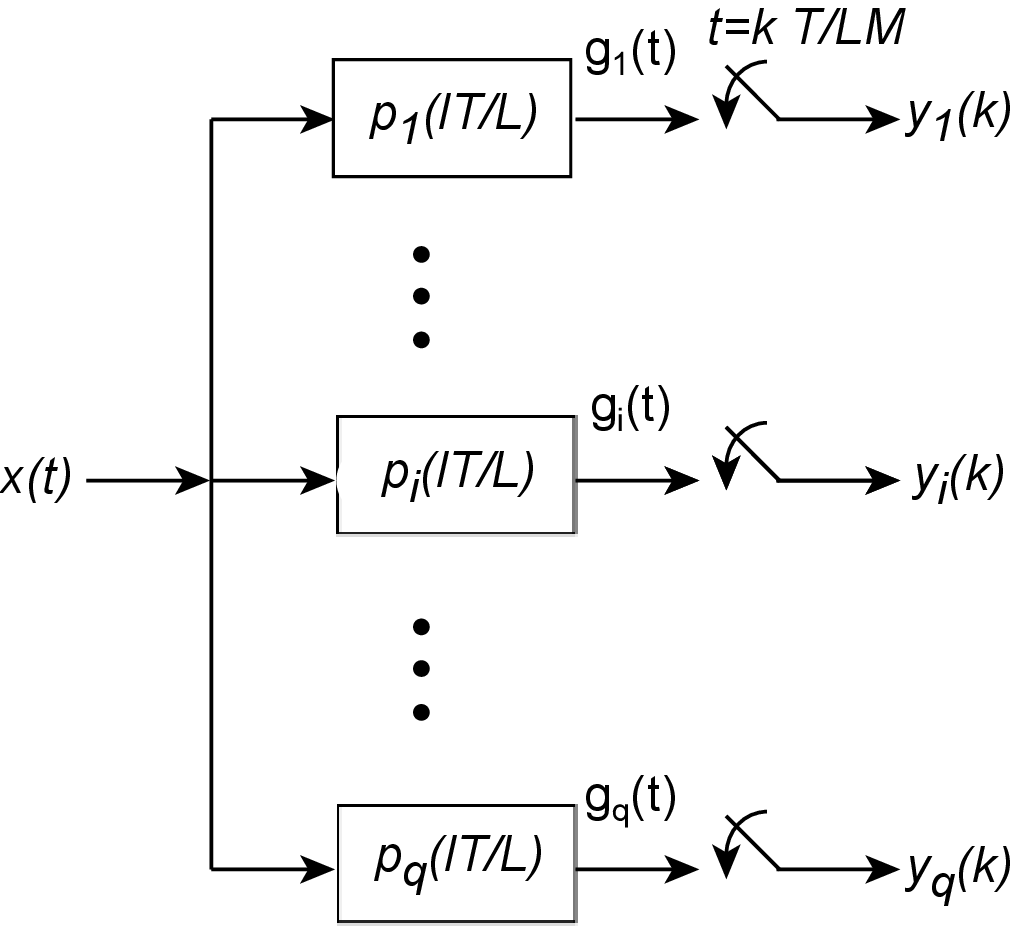}}
\end{minipage}\label{subfig:timeblock_structure}}
\subfigure[ ]{\begin{minipage}{.48\textwidth}
\centerline{\includegraphics[width=7cm]{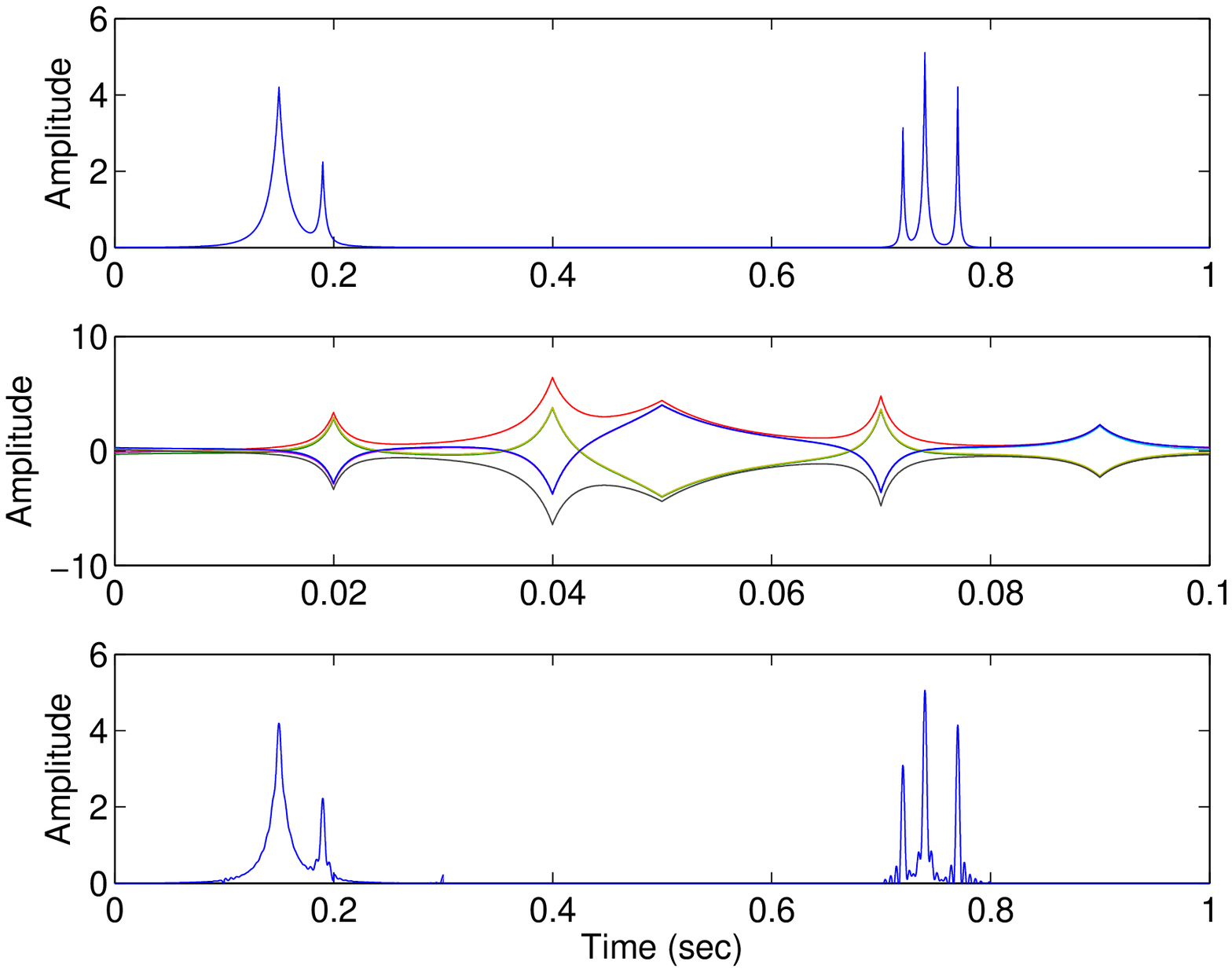}}
\end{minipage}\label{subfig:timeblock_example}}
\caption{\small (a) Schematic diagram of a continuous-time block-sparse sampling system. 
The system is a generalization of random convolution as originally proposed in~\cite{romberg2009}.
Each channel convolves $x(t)$ with a random sequence and then samples the result at a low rate.
(b) Top panel: Simulated block sparse time domain signal on the unit interval.
The signal is a modified version of the ``bumps'' test signal from the WaveLab toolbox~\cite{wavelab}.
Middle panel: Overlay plot of the signals $g_{i}(t)$ resulting from the random convolution.
Here, $x(t)$ was sampled with a 8 channel system.
Bottom panel: Reconstructed signal approximation.
Each $0.1$ second duration segment is a linear approximation (Fourier basis) of $x(t)$ on that segment (normalized squared error$=0.055$).}
\end{figure}

The proposed system combines block convolution with the MWC architecture and the random convolution ideas of Romberg to obtain a sampling system for continuous-time block-sparse signals (see Figure~\ref{subfig:timeblock_structure}).
The resulting system can also be interpreted as the time domain analogue of the MWC.

Matusiak and Eldar~\cite{matusiak_eldar2011} recently proposed a sampling system targeting a nearly identical signal class.
Their system also shares structural similarities to the MWC but leverages Gabor frames instead of block convolution.  

\textbf{Sampling System.} Let $x(t)$ be a continuous block-sparse signal on the interval $[0,T]$ and let $\B{Z}$ be a Bernoulli random variable taking values $\pm1$ with equal probability.
Denote by $\{p_{i}(l), l=0,\dotsc,L-1\}_{i=0}^{q-1}$ an ensemble of random vectors drawn from $\B{Z}$ ($L,q\in\mathbb{Z}^{+}$). 
The sampling system has $q$ channels and operates in parallel like the MWC: the $i^{\text{th}}$ channel convolves $x(t)$ with $p_{i}(l)$, resulting in the continuous-time signal
\begin{equation}\label{equ:block_convolution_a}
g_{i}(t)=\sum_{l=0}^{L-1} p_{i}(l) x(t-l\tfrac{T}{L}),
\end{equation}
and then uniformly samples $g_{i}(t)$.
Note that by construction the convolution in~\eqref{equ:block_convolution_a} is a block-convolution (see Section~\ref{subsect:MWCmultitone}) and is the standard filtering operation underlying waveform synthesis~\cite[p. 123]{roberts_mullisDSP}.
Restricting the time axis to the interval $[0,T/L]$, \eqref{equ:block_convolution_a} becomes
\begin{equation}\label{equ:block_convolution_b}
g_{i}(t)\rect{\tfrac{2L}{T}(t-\tfrac{T}{2L})}=\sum_{l=0}^{L-1} p_{i}(l) x(t-l\tfrac{T}{L})\rect{\tfrac{2L}{T}(t-\tfrac{T}{2L})},
\end{equation}
for $i=0,\ldots,q-1$.
This system of equations is similar to the frequency domain description of the MWC (equation~\eqref{equ:mwc_freqdomain}) in that the unknowns are segments of the continuous signal of interest; here we are interested in recovering $\bigl \{x(t-l\tfrac{T}{L})\rect{\tfrac{2L}{T}(t-\tfrac{T}{2L})} \bigr\}_{l}$ whereas for the MWC we want to recover $\bigl \{X(\omega-m\tfrac{2\pi\mwcW}{\mwcL})\rect{\tfrac{\mwcL}{\pi \mwcW}\omega}\bigr \}_{m}$.
Consequently, given the block sparsity of $x(t)$, one could proceed to solve the linear system in a manner similar to that proposed by Mishali and Eldar for the MWC.
Alternatively, one could simply discretize the time axis, form a MMV problem from~\eqref{equ:block_convolution_b} and solve for samples of the segments $\bigl \{x(t-l\tfrac{T}{L})\rect{\tfrac{2L}{T}(t-\tfrac{T}{2L})}\bigl \}_{l}$.
(This type of approach was described in Section~\ref{subsect:rd_with_multiband}.)
Either way, the solution that is obtained would be an approximation of the original block-sparse time domain signal:
the MWC reconstruction method would yield a linear approximation (see Section~\ref{subsect:signal_models_mwc}), and discretizing the time axis would produce samples between which one would have to interpolate.
Below, we present another method to compute a linear approximation that, unlike the Mishali and Eldar method, uses CS techniques to directly retrieve both the support of $x(t)$ (at resolution $L$) and the coefficients of the approximation simultaneously.

Representing the segments $\bigl \{x(t-l\tfrac{T}{L})\rect{\tfrac{2L}{T}(t-\tfrac{T}{2L})} \bigr\}_{l}$ in an appropriate orthogonal basis,~\eqref{equ:block_convolution_b} may be written as
\begin{equation}\label{equ:block_convolution_c}
g_{i}(t)\rect{\tfrac{2L}{T}(t-\tfrac{T}{2L})} =\sum_{l=0}^{L-1} p_{i}(l) \sum_{n=0}^{\infty}  \alpha_{l}(n)\psi_{n}(t),
\end{equation}
where $\alpha_{l}(n)=\big\langle x(t-l\tfrac{T}{L})\rect{\tfrac{2L}{T}(t-\tfrac{T}{2L})},\psi_{n}(t) \big\rangle$.
Sampling at a rate $\tfrac{LM}{T}$ Hz over the interval $[0,\tfrac{T}{L}]$ yields
\begin{align}
y_{i}(k)&=g_{i}(k\tfrac{T}{LM}) \nonumber\\
&=\sum_{l=0}^{L-1}p_{i}(l) \sum_{n=0}^{\infty}  \alpha_{l}(n)\psi_{n}(k\tfrac{T}{LM}),  \label{equ:block_convolution_d}
\end{align}
for $k=0,\dotsc,M-1$ and $i=0,\dotsc q-1$.
Thus to recover a $D$-term linear approximation, one needs to solve for the coefficient matrix $\B{A}$ in the matrix form of~\eqref{equ:block_convolution_d},
\begin{equation}\label{equ:block_convolution_e}
\B{Y}=\B{\Phi}\B{A}\B{\Psi}
\end{equation}
where
\begin{align*}
&\B{Y}_{i,k}=y_{i}(k) \\
&\B{\Phi}_{i,l}=p_{i}(l) \\
&\B{A}_{l,n}=\alpha_{l}(n) \\
&\B{\Psi}_{n,k}=\psi_{n}(k\tfrac{T}{LM}) 
\end{align*}
for $i=0,\dotsc,q-1$, $k=0,\dotsc,M-1$, $l=0,\dotsc,L-1$, and $n=0,\dotsc,D-1$.
Given the samples $\{y_{i}(k)\}$ and the matrices $\B{\Phi}$ and $\B{\Psi}$, one can form an MMV problem by post multiplying both sides of~\eqref{equ:block_convolution_e} by the right-inverse of $\B{\Psi}$, when it exists.
This yields
\begin{equation}\label{equ:block_convolution_f}
\B{Y}\B{\Psi}^{\dagger}=\B{\Phi}\B{A}
\end{equation}
where $\B{\Psi}^{\dagger}$ denotes the right-inverse of $\B{\Psi}$.
The right-inverse exists if and only if the columns of $\B{\Psi}$ span $\real^{D}$~\cite{strang1988}, which is only possible if $D\leq M$, or equivalently, when the approximation order is less than or equal to the number of acquired samples.
Thus, assuming $\B{\Psi}^{\dagger}$ exists, the maximum approximation order $D$ that can be recovered with this scheme equals the number of samples $M$ acquired per channel. 
Conversely, the minimum number of samples needed to successfully recover a $D$-order approximation is $M$.
However, the desire to recover an approximation that is as accurate as possible while using a minimum number of samples suggests that setting $D=M$ is an optimal choice.
Thus, in what follows, we assume $\B{\Psi}$ is a \emph{square} matrix that has an inverse $\B{\Psi}^{-1}$.
The linear system in~\eqref{equ:block_convolution_f} therefore becomes
\begin{equation}\label{equ:block_convolution_g}
\B{Y}\B{\Psi}^{-1}=\B{\Phi}\B{A}.
\end{equation}

From a CS perspective, $\B{\Phi}$ is a $q\times L$ Bernoulli sensing matrix where $q<L$, and the elements of the product $\B{Y}\B{\Psi}^{-1}$ are the CS measurements.
In this case, and in contrast to the RD and the MWC, these measurements are not simply the samples acquired by the system (c.f. equations~\eqref{equ:rd_matrixform} and~\eqref{equ:mwc_imv}), but are linear combinations of the samples.
The coefficient matrix $\B{A}$ can be solved for using any existing MMV CS solver, see e.g.~\cite{cotter_etal2005,greedy_pursuit_tropp_etal2006,convex_relax_tropp_etal2006,chen_huo2006,mishali_eldar2008,davies_eldar2010}.
The conditions defining when a unique solution exists has been extensively studied and depends on the properties of $\B{\Phi}$ and the degree to which $\B{A}$ is sparse (again, see~\cite{cotter_etal2005,greedy_pursuit_tropp_etal2006,convex_relax_tropp_etal2006,chen_huo2006,mishali_eldar2008,davies_eldar2010}).
Note that because $x(t)$ is assumed to be block-sparse, it follows that $\B{A}$ is a joint sparse matrix.

Letting $\widehat{\B{A}}_{l,n}=\widehat{\alpha}_{l}(n)$ denote the entries of the CS solution of~\eqref{equ:block_convolution_g}, $x(t)$ can be approximated (reconstructed) by computing an approximation for each segment $\bigl \{x(t-l\tfrac{T}{L})\rect{\tfrac{2L}{T}(t-\tfrac{T}{2L})} \bigr\}_{l}, l=0,\dotsc,L-1$, 
\begin{equation*}
x(t-l\tfrac{T}{L})\rect{\tfrac{2L}{T}(t-\tfrac{T}{2L})} \approx \sum_{n=0}^{D-1} \widehat{\alpha}_{l}(n) \psi_{n}(t)
\end{equation*}
and then concatenating the $L$ segment approximations.

\textbf{Example.}
Consider the continuous-time block sparse signal depicted in the top panel of Figure~\ref{subfig:timeblock_example}.
Samples of the filtered signals $g_{i}(t)$ (middle panel) are acquired by an 8 channel system ($q=8$) with a sampling rate of $400$ Hz ($M=40, LM/T=400$) where the time axis is partitioned into 10 segments ($L=10$).
Using a Fourier basis, $\psi_{n}(t)=e^{j\tfrac{2\pi}{T/L}tn}$, the linear system~\eqref{equ:block_convolution_e} was solved using S-OMP~\cite{greedy_pursuit_tropp_etal2006} resulting in $L$ linear approximations ($D=40$) of the segments $\bigl \{x(t-l\tfrac{T}{L})\rect{\tfrac{2L}{T}(t-\tfrac{T}{2L})} \bigr \}_{l}$.
The reconstructed signal is shown in the bottom panel of Figure~\ref{subfig:timeblock_example}.
In this case, the reconstructed signal is a faithful representation of $x(t)$ (normalized squared error $=0.055$).
Clearly, the quality of the reconstruction depends on the accuracy of the approximations and on how well a particular CS algorithm solves~\eqref{equ:block_convolution_g}.
For a given application, some bases will be more appropriate than others. 
For example, if $x(t)$ contains discontinuities, a wavelet basis would likely provide a better approximation than a Fourier basis.

Strictly speaking continuous-time block sparse signals are not bandlimited, but if one examines the spectrum of the test signal in Figure~\ref{subfig:timeblock_example}, one would discover that the signal is ``essentially'' bandlimited to about 1000 Hz. 
Thus, it could be argued according to the Shannon-Nyquist sampling theorem that a sampling rate of about 2000 Hz would be required to accurately capture this signal. 
Relative to 2000 Hz, 400 Hz represents a five fold reduction in sampling rate. 

\section{Conclusion}\label{sect:conclusion}
In this paper, we showed that the sampling mechanisms of the RD and the MWC can both be thought of as being based on the underlying concept of random filtering or random convolution.
The most substantial difference between the systems stems from the specific form of their sampling functions (or random filters) and from the assumed signal models.
The RD has sampling functions that have finite temporal extent but infinite spectral support; the MWC employs sampling functions that have finite spectral support but infinite temporal support.
The randomness in the sampling functions is a hallmark of CS theory that is fundamental in guaranteeing unique solutions to the underdetermined linear systems that characterise the RD and the MWC.

Block convolution is also an important property that differentiates the MWC from the RD because it is one approach that effectively processes infinite dimensional signals that have a block structure.
The absence of this property is one primary reason the RD cannot, in general, reconstruct multiband signals.
We incorporated block convolution into a new sampling system that samples continuous-time block-sparse signals. 

We also offered two novel insights into how the MWC behaves with regard to the underlying sparsity assumption.
We showed that if the number of channels is near the minimum required, relatively small changes in the number of channels (or equivalently if the sparsity changes for a given number of channels) can cause significant reconstruction errors.
On the other hand, we provided evidence that the MWC is robust to increases in the width of the spectral bands caused by windowing.  

From this paper's perspective, one can begin to consider generalisations to the RD and the MWC that target different signal classes, in particular, CS  sampling systems that have different time-frequency characterisations.
For example, a system that ``compressively'' samples radar pulses and chirps in an efficient time-frequency manner could possibly offer a means to effectively detect and classify these signals while avoiding the overhead of sampling several bands simultaneously or reconstructing the Nyquist equivalent signal.
This paper takes a step towards this goal by reconciling some of the core ideas behind these sampling systems.

\section{Appendices}
The following analyses yield basic time and frequency domain descriptions of the sampling strategies.
We employ standard Fourier transform properties without explicit explanation for the sake of conciseness.
The notational style is that of~\cite{roberts_mullisDSP}.
To denote transform pairs, we use the shorthand notation,
\begin{equation*}
x(t)\xleftrightarrow{\hspace*{7pt}\text{FT\hspace*{7pt}}} X(\omega),
\end{equation*}
and use the abbreviations FT, FS, DTFT, and DFT when referring to the Fourier transform, the Fourier series, the discrete time Fourier transform, and the discrete Fourier transform, respectively.
Also recall the definitions: $\sinc{(x)}=\sin{(x)}/x, \,x\in \real$ and
\begin{equation*}
\text{rect}(x)=\begin{cases}
1\quad\text{for~}-1\leq x \leq 1\\
0\quad\text{otherwise}\end{cases}.
\end{equation*}

\subsection{Derivations of Equations~\eqref{equ:rd_linearsystem} and~\eqref{equ:rd_freqdomain}} \label{app:RD}
\textbf{Time domain description.}
Let $x(t)$ be a sparse multitone signal on $[0,T]$ and recall the following transform pairs from Section~\ref{subsect:sampling_with_rd}:
\begin{align*}
x(t) &\xleftrightarrow{\hspace*{3pt}\text{FS;1/T\hspace*{3pt}}} X(n) \\
p(t) &\xleftrightarrow{\hspace*{3pt}\text{FS;1/T\hspace*{3pt}}} P(n) \\
h(t)=\text{rect}\bigl(\tfrac{2M}{T}t-1\bigr)& \xleftrightarrow{\hspace*{5pt}\text{FT\hspace*{5pt}}} H(\omega)=\tfrac{T}{M}\sinc{(\tfrac{T}{2M}\omega)}e^{-j\omega}.
\end{align*}
By inspection of Figure~\ref{subfig:rd}, the time domain description of the RD is
\begin{align*}                                                                                               
g(t)&=x(t)p(t)*h(t)\\
&= \int_{0}^{T}x(\tau)p(\tau)h(t-\tau)~d\tau \\
&=\int_{t-\tfrac{T}{M}}^{t}x(\tau)p(\tau)~d\tau \\
&=\sum_{n=-N/2}^{N/2-1} X(n) \int_{t-\tfrac{T}{M}}^{t}p(\tau) e^{j\frac{2\pi}{T}n\tau}~d\tau,
\end{align*}
where $*$ denotes convolution.
Sampling at $t=(k+1)\tfrac{T}{M}$ for $k=0, 1,\dotsc, M-1$ yields
\begin{align}
y(k)&=g\bigl((k+1)\tfrac{T}{M}\bigr) \nonumber \\
&=\sum_{n=-N/2}^{N/2-1} X(n) \int_{k\tfrac{T}{M}}^{(k+1)\tfrac{T}{M}} p(\tau)e^{j\frac{2\pi}{T}n\tau}~d\tau \nonumber \\
&=\sum_{n=-N/2}^{N/2-1} X(n) \sum_{m=0}^{N/M-1} \int_{k\tfrac{T}{M}+\tfrac{m}{W}}^{k\tfrac{T}{M}+\tfrac{m+1}{W}}  p(\tau) e^{j\frac{2\pi}{T}n\tau}~d\tau \nonumber\\
&=\sum_{n=-N/2}^{N/2-1} \sum_{m=0}^{N/M-1} p(k\tfrac{T}{M}+\tfrac{m}{W}) X(n) \int_{k\tfrac{T}{M}+\tfrac{m}{W}}^{k\tfrac{T}{M}+\tfrac{m+1}{W}}  e^{j\frac{2\pi}{T}n\tau}~d\tau \label{equ:rd_a} 
\end{align}
where each step follows from the additivity of the integral and the specific nature of $p(t)$.
Evaluating the integral, we obtain
\begin{equation}\label{equ:rd_c}
\int_{k\tfrac{T}{M}+\tfrac{m}{W}}^{k\tfrac{T}{M}+\tfrac{m+1}{W}}  e^{j\frac{2\pi}{T}n\tau}~d\tau = \begin{cases} T \,\dfrac{e^{j\frac{2\pi}{N}n}-1}{j2\pi n} \,e^{j\tfrac{2\pi}{N}n\bigl(k\tfrac{N}{M}+m\bigr)}  &n\neq 0 \\
\frac{1}{W}  &n= 0
\end{cases}
\end{equation}
Substituting~\eqref{equ:rd_c} into~\eqref{equ:rd_a} and letting $l=k\tfrac{N}{M}+m$,~\eqref{equ:rd_a} may be rewritten as
\begin{equation*}
y(k)=  \sum_{n=-N/2}^{N/2-1} \alpha(n) \,X(n) \sum_{l=k\tfrac{N}{M}}^{(k+1)\tfrac{N}{M}-1} p_{l} \, e^{j\tfrac{2\pi}{N}nl},
\end{equation*}
for $k=0,\dotsc M-1$ where $p_{l}=p(l/W)$ and 
\begin{equation*}
\alpha(n)=\begin{cases} T \dfrac{e^{j\frac{2\pi}{N}n}-1}{j2\pi n} & n\neq 0 \\
1/W  & n=0 \end{cases}
\end{equation*}  

\vspace*{7pt}
\noindent\textbf{Frequency domain description.}
We also have the following frequency domain description of the RD.

\vspace*{7pt}
\noindent \textbf{Multiplication/Convolution:}
\begin{equation*}
x(t)p(t)\xleftrightarrow{\hspace*{3pt}\text{FS;1/T}\hspace*{3pt}} \sum_{m=-\lfloor n-\frac{N}{2}\rfloor+1}^{\lfloor n-\frac{N}{2}\rfloor} P(m) X(n-m) 
\end{equation*}

\vspace*{7pt}
\noindent \textbf{Convolution (filtering)/Multiplication:}
\begin{align*}
&g(t)=x(t)p(t)*h(t) \\ 
& \updownarrow{\hspace*{3pt}\text{\scriptsize{FS; 1/T}}} \\
&G(n)= \sum_{m=-\lfloor n-\frac{N}{2}\rfloor+1}^{\lfloor n-\frac{N}{2}\rfloor} P(m) X(n-m)H(j\tfrac{2\pi}{T}n) \\
&~\qquad =\frac{T}{M}\sum_{m=-\lfloor n-\frac{N}{2}\rfloor+1}^{\lfloor n-\frac{N}{2}\rfloor} P(m) X(n-m) e^{-j\frac{2\pi}{T}n} \sinc{(\tfrac{\pi}{M}n)}
\end{align*}

\vspace*{7pt}
\noindent \textbf{Sampling/Aliasing:}
\begin{equation*}
y(k)=g\bigl((k+1)\tfrac{T}{M}\bigr) \xleftrightarrow{\hspace*{3pt}\text{DFT;M}\hspace*{3pt}} Y(n)=M \sum_{l=-\infty}^{\infty} G(n-lM)
\end{equation*}
Because $Y(n)$ is $M$-periodic, we can, without loss of information, restrict it to one period.
This means we need only consider one term in the summation over $l$.
Retaining the $l=0$ term yields
\begin{equation*}
Y(n)= T \sum_{m=-\infty}^{\infty} P(m) e^{-j\frac{2\pi}{T}n}\sinc{(\tfrac{\pi}{M}n)}  \,X(n-m),
\end{equation*}
for $n=0,\dotsc,M-1$.

\subsection{Derivation of Equation~\eqref{equ:mwc_freqdomain_a}} \label{app:MWC}
Let $x(t)$ be a sparse multiband signal and recall the following transform pairs from Section~\ref{subsect:mwc}:
\begin{align*}
x(t) &\xleftrightarrow{\hspace*{5pt}\text{FT\hspace*{5pt}}} X(\omega) \\
p_{i}(t) &\xleftrightarrow{\hspace*{3pt}\text{FS;}\mwcW/\mwcL \hspace*{3pt}}P_{i}(m) \\
h(t)=\tfrac{\pi \mwcW}{\mwcL}\sinc(\tfrac{\pi \mwcW}{\mwcL}t)& \xleftrightarrow{\hspace*{5pt}\text{FT\hspace*{5pt}}} H(\omega)=\rect{\tfrac{\omega\mwcL}{\pi \mwcW}}.
\end{align*}
We then have the following time and frequency domain descriptions for the $i^{\text{th}}$ channel, $i=0,\dotsc,\mwcq-1$ of the MWC.

\vspace*{7pt}
\noindent \textbf{Multiplication/Convolution:}
\begin{align*}
x(t)p_{i}(t)\xleftrightarrow{\hspace*{5pt}\text{FT}\hspace*{5pt}} &\sum_{m=-\infty}^{\infty} P_{i}(m) X(\omega-m\omega_{p}) \\
&\quad =\sum_{m=\lceil(\omega-\pi \mwcW)/\omega_{p}\rceil+1}^{\lfloor(\omega+\pi \mwcW)/\omega_{p}\rfloor} P_{i}(m) X(\omega-m\omega_{p})
\end{align*}
where $\omega_{p}=2\pi \mwcW/\mwcL$ radians per second.
The summation limits are finite for a given $\omega$ because $x(t)$ is assumed to be bandlimited.

\vspace*{7pt}
\noindent \textbf{Convolution (filtering)/Multiplication:}
\begin{align*}
&g_{i}(t)=x(t)p_{i}(t)*h(t) \\
& \updownarrow{\hspace*{3pt}\text{\scriptsize{FT}}} \\
&G_{i}(\omega)= \sum_{m=\lceil(\omega-\pi \mwcW)/\omega_{p}\rceil+1}^{\lfloor(\omega+\pi \mwcW)/\omega_{p}\rfloor} P_{i}(m) X(\omega-m\omega_{p})H(\omega) \\
&~ ~\qquad =\sum_{m=-\lfloor\frac{1}{2}+\frac{\pi \mwcW}{\omega_p}\rfloor+1}^{\lfloor\frac{1}{2}+\frac{\pi \mwcW}{\omega_p}\rfloor} P_{i}(m) X(\omega-m\omega_{p}) \rect{\tfrac{2\omega}{\omega_p}}
\end{align*}
Note that the low pass filter windows $X(\omega)$ and its translates (i.e., restricts them to the interval $[-\omega_p/2,\omega_p/2]$) and hence removes the dependence on $\omega$ in the summation limits. 

\vspace*{7pt}
\noindent \textbf{Sampling/Aliasing:}
\begin{align}
&y_{i}(k)=g_{i}(kT_{s}) \xleftrightarrow{\hspace*{5pt}\text{DTFT}\hspace*{5pt}}
Y_{i}(e^{j\omega\frac{\mwcM}{\mwcW}})=\frac{\mwcW}{\mwcM}\sum_{n=-\infty}^{\infty} G_{i}(\omega+n\omega_{s}) \label{equ:mwc_f}
\end{align}
Observe that the translates of $G_i(\omega)$ in~\eqref{equ:mwc_f} do not overlap because $G_i(\omega)$ is bandlimited to $[-\omega_p/2,\omega_p/2]$ and by assumption $\omega_p \leq \omega_s$ (see Section~\ref{subsect:mwc}). 
We can therefore, without loss of information, restrict $Y_{i}(e^{j\omega\frac{\mwcM}{\mwcW}})$ to one period \big($Y_{i}(e^{j\omega\frac{\mwcM}{\mwcW}})$ is $2\pi \mwcW/\mwcM$-periodic\big). 
This means we need only consider one term in the summation over $n$ in~\eqref{equ:mwc_f}.
We choose to retain the $n=0$ term and thus have the DTFT
\begin{equation*}
\begin{split}
Y_{i}(e^{j\omega\frac{\mwcM}{\mwcW}})&\rect{\tfrac{\mwcM}{\pi\mwcW}\omega} =\\
&\frac{\mwcW}{\mwcM} \sum_{m=-\lfloor\frac{1}{2}(\mwcL+1)\rfloor+1}^{\lfloor\frac{1}{2}(\mwcL+1)\rfloor}  P_{i}(m) X(\omega-m\omega_{p}) \rect{\tfrac{2\omega}{\omega_p}}.
\end{split}
\end{equation*}
The Fourier series coefficients of $p_{i}(t)$ can now be directly computed,
\begin{align}
P_{i}(m)&= \frac{1}{T_P} \int_{0}^{T_p} p_{i}(t)~e^{-j\frac{2\pi}{T_p}mt}~dt \nonumber \\
&= \frac{1}{T_P} \sum_{l=0}^{\mwcL-1} \int_{l\frac{T_p}{\mwcL}}^{(l+1)\frac{T_p}{\mwcL}} p_{il}~e^{-j\frac{2\pi}{T_p}mt}~dt \nonumber \\
&= \begin{cases}\sum_{l=0}^{\mwcL-1} \frac{p_{il}}{j2\pi m} \bigl(1-e^{-j\frac{2\pi}{\mwcL}m}\bigr) e^{-j\frac{2\pi}{\mwcL}ml}, & m\neq 0 \\
\tfrac{1}{\mwcL}\sum_{l=0}^{\mwcL-1}p_{il},& m=0,\end{cases} \label{equ:mwc_g}
\end{align}
where $p_{il}=p_{i}(t)$ for $t\in[l/\mwcW,(l+1)/\mwcW)$, to obtain
\begin{equation*}
\begin{split}
&Y_{i}(e^{j\omega\frac{\mwcM}{\mwcW}})\rect{\tfrac{\mwcM}{\pi\mwcW}\omega} =\\
&\sum_{m=-\lfloor\frac{1}{2}(\mwcL+1)\rfloor+1}^{\lfloor\frac{1}{2}(\mwcL+1)\rfloor} \beta(m)\,X(\omega-m \tfrac{2 \pi\mwcW}{\mwcL})\rect{\tfrac{\mwcL}{\pi \mwcW}\omega} \sum_{l=0}^{\mwcL-1} p_{il}\,   e^{-j\frac{2\pi}{\mwcL}ml},
\end{split}
\end{equation*}
where
\begin{equation*}
\beta(m)=\begin{cases} \dfrac{\mwcW}{\mwcM} \dfrac{1-e^{-j\frac{2\pi}{\mwcL}m}}{j2\pi m}, & m\neq 0 \\
1/\mwcL, & m=0 \end{cases}.
\end{equation*}

\subsection{Derivation of Equation~\eqref{equ:mwc_tone_b}} \label{app:MWCmultitone}
Let $x(t)$ be a sparse multitone signal on $[0,T]$ and recall the following transform pairs and parameter relations from Section~\ref{subsect:MWCmultitone}:
\begin{align*}
x(t) &\xleftrightarrow{\hspace*{3pt}\text{FS;1/T\hspace*{3pt}}} X(n) \\
p_{i}(t) &\xleftrightarrow{\hspace*{3pt}\text{FS;}1/T_p \hspace*{3pt}}P_{i}(m) \\
h(t)=\tfrac{\pi W}{\mwcL}\sinc(\tfrac{\pi W}{\mwcL}t)& \xleftrightarrow{\hspace*{5pt}\text{FT\hspace*{5pt}}} H(\omega)=\rect{\tfrac{\mwcL}{\pi W}\omega} \\
T=\frac{N\mwcL}{W}, ~N\in\mathbb{Z},&~N>1,~ T_p=\frac{\mwcL}{W},~ T_s=\frac{\mwcM}{W}
\end{align*}
Also recall the simplifying assumption $\mwcL=\mwcM$.
In the following analysis, we use the periodic extension of $x(t)$, denoted by $\tilde{x}(t)$, because it is defined for all $t\in\real$ (thereby simplifying calculations) and has the same FS coefficients as $x(t)$.
Its use does not imply that $x(t)$ must be replicated before the MWC samples it.
Rather it implies a discretisation step of the frequency axis that must otherwise be explicitly carried out to form the MMV problem in~\eqref{equ:mwc_multitone_mmv}.

\noindent \textbf{Multiplication/Convolution:}
\begin{equation*}
\tilde{x}(t)p_{i}(t)\xleftrightarrow{\hspace*{3pt}\text{FS;$1/T$}\hspace*{3pt}} 
\sum_{m=\lceil \frac{1}{N}(n-\frac{N\mwcL}{2})\rceil+1}^{\lfloor\frac{1}{N}(n+\frac{N\mwcL}{2})\rfloor} P_{i}(m) X(n-Nm)
\end{equation*}
Note that the left hand side is defined for all $t\in\real$ and the right is defined for all $n\in\mathbb{Z}$. 
The summation limits are finite (for a given $n$) because the harmonics of $x(t)$ are assumed to be bounded.

\vspace*{7pt}
\noindent \textbf{Convolution (filtering)/Multiplication:}
\begin{align*}
&g_{i}(t)=\tilde{x}(t)p_{i}(t)*h(t) \\ 
& \updownarrow{\hspace*{3pt}\text{\scriptsize{FS;1/T}}} \\
&G_{i}(n)= \sum_{m=\lceil \frac{1}{N}(n-\frac{N\mwcL}{2})\rceil+1}^{\lfloor\frac{1}{N}(n+\frac{N\mwcL}{2})\rfloor} P_{i}(m)\, X(n-Nm) H\bigl(\tfrac{2\pi}{T}n\bigr), \\
&~~\qquad=\sum_{m=-\lfloor\frac{1}{2}(\mwcL+1)\rfloor+1}^{\lfloor\frac{1}{2}(\mwcL+1)\rfloor} P_{i}(m)\, X(n-Nm) \rect{\tfrac{2}{N}n}
\end{align*}
where we note that $g_{i}(t)$ is $1/T$-periodic and $H\bigl(\tfrac{2\pi}{T}n\bigr)=H(\omega)\vert_{\omega=2\pi n/T}$.
Note that the low pass filter windows $X(n)$ and its translates, i.e., restricts them to the interval $[-\tfrac{\pi W}{\mwcL},\tfrac{\pi W}{\mwcL}]$ (or $[-N/2,N/2]$) and hence removes the dependence on $n$ in the summation limits. 

\vspace*{7pt}
\noindent \textbf{Sampling/Aliasing:}
\begin{equation}\label{equ:MWCmultitone_a}
y_{i}(k)=g_{i}(kT_{s}) \xleftrightarrow{\hspace*{3pt}\text{DFT;$N$}\hspace*{3pt}} Y_{i}(n)=\frac{1}{N} \sum_{l=-\infty}^{\infty} G_{i}\bigl(n-lN\bigr)
\end{equation}
Because $G_{i}(n)$ is ``bandlimited'' to $[-N/2,N/2]$, its translates in~\eqref{equ:MWCmultitone_a} do not overlap.
Consequently, it is sufficient to only consider one period of $Y_{i}(n)$, or equivalently, only one term in the above summation.
Choosing the $l=0$ term,~\eqref{equ:MWCmultitone_a} becomes
\begin{equation}\label{equ:MWCmultitone_b}
Y_{i}(n)=\frac{1}{N} \sum_{m=-\lfloor\frac{1}{2}(\mwcL+1)\rfloor+1}^{\lfloor\frac{1}{2}(\mwcL+1)\rfloor} P_{i}(m)\, X(n-Nm) \rect{\tfrac{2}{N}n}, 
\end{equation}
for $ -N/2\leq n < N/2$.
By substituting the expression for $P_{i}(m)$ from~\eqref{equ:mwc_g} into~\eqref{equ:MWCmultitone_b}, we have the relation
\begin{equation*}
Y_{i}(n)=\sum_{m=-\lfloor\frac{1}{2}(\mwcL+1)\rfloor+1}^{\lfloor\frac{1}{2}(\mwcL+1)\rfloor} \eta(m) \,X(n-Nm)\rect{\tfrac{2}{N}n} \sum_{l=0}^{\mwcL-1} p_{il} \, e^{-j\frac{2\pi}{\mwcL}lm},
\end{equation*}
where $p_{il}=p_{i}(t)$ for $t\in[l/W,(l+1)/W)$ and
\begin{equation*}
\eta(m)=\begin{cases}
\dfrac{1}{N}\dfrac{1-e^{-j\frac{2\pi}{\mwcL}m}}{j2\pi m} & m\neq 0\\
1/\mwcL & m=0
\end{cases}.
\end{equation*}

\single\small
\bibliographystyle{IEEEbib}
\bibliography{lexa}

\end{document}